\documentclass[aps,groupedaddress,twocolumn,notitlepage,superscriptaddress,10pt]{revtex4-1}

\usepackage[cp1252]{inputenc}

\usepackage{float}
\usepackage{graphicx}  
\usepackage{xcolor}
\usepackage{dcolumn}   
\usepackage{bm}        
\usepackage{braket}
\usepackage{amssymb}   
\usepackage{epstopdf}
\usepackage{xfrac}
\usepackage{subfigure}
\usepackage{anyfontsize}

\usepackage{amsmath}
\usepackage{amsthm}
\usepackage{amsfonts}
\usepackage{bbm}
\definecolor{darkblue}{rgb}{0.1,0.2,0.6}
\definecolor{darkred}{rgb}{0.8,0.1,0.2}
\usepackage[colorlinks,citecolor=darkblue,linkcolor=darkblue,urlcolor=darkblue]{hyperref} 
\usepackage{enumerate}
\usepackage{setspace}
\usepackage{url}  
\usepackage{mathrsfs}

\theoremstyle{plain}
\newtheorem{thm}{Theorem}

\newtheorem{prop}{Proposition}
\newtheorem{cor}{Corollary}
\newtheorem{chk}{Check}
\newtheorem{rmk}{Remark}
\newtheorem{conj}{Conjecture}

\theoremstyle{definition}

\usepackage{tensor} 

\newcommand{\Hi}{\mathcal{H}}
\newcommand{\sT}{\sf T}
\newcommand{\sA}{\sf A}

\newcommand{\Z}{\mathbb{Z}}

\newcommand{\I}{\mathbb{I}}

\newcommand{\Tr}{\text{Tr}}
\newcommand{\tr}{\text{Tr}}

\newcommand{\norm}[1]{\left\lVert#1\right\rVert}

\newcommand{\vac}{\ket{0}\bra{0}}

\begin{document}
\title{Entanglement negativity of fermions: monotonicity, separability criterion, and classification of few-mode states}
\author{Hassan Shapourian}
\affiliation{James Franck Institute and Kadanoff Center for Theoretical Physics, University of Chicago, IL 60637, USA.}
\affiliation{Kavli Institute for Theoretical Physics, University of California, Santa Barbara, CA 93106, USA.}
\author{Shinsei Ryu}
\affiliation{James Franck Institute and Kadanoff Center for Theoretical Physics, University of Chicago, IL 60637, USA.}

\begin{abstract}

We study quantum information aspects of the fermionic entanglement negativity recently introduced in [\href{https://journals.aps.org/prb/abstract/10.1103/PhysRevB.95.165101}{Phys. Rev. B {\bf 95}, 165101 (2017)}] based on the fermionic partial transpose.
In particular, we show that it is an entanglement monotone under the action of local quantum operations and classical communications--which preserves the local fermion-number parity-- and satisfies other common properties expected for an entanglement measure of mixed states. 
We present fermionic analogs of tripartite entangled states such as W and Greenberger-Horne-Zeilinger states and compare the results of bosonic and fermionic partial transpose in various fermionic states, where we explain why the bosonic partial transpose fails in distinguishing separable states of fermions. 
Finally,  we explore a set of entanglement quantities which distinguish different classes of entangled states of a system with two and three fermionic modes. 
In doing so, we prove that vanishing entanglement negativity  is a necessary and sufficient condition for separability of $N\geq 2$ fermionic modes  with respect to the bipartition into one mode and the rest.
We further conjecture that the entanglement negativity of inseparable states which mix local fermion-number parity is always non-vanishing.
\end{abstract}

\maketitle


\section{Introduction}

Quantum entanglement is a unique property of quantum systems that has no classical analog. 
Being intrinsic and fundamental, quantum entanglement is a frontier in various research areas ranging from
many-body physics to spacetime physics and quantum information.
A fundamental question is how the entanglement between two parts of a quantum system can be quantified in terms of some computable measures. 

In general, a quantum system can be in a pure state or a mixed state. 
Pure states are described by wave functions, for example, the zero temperature
ground state wave function of a Hamiltonian.
Bipartite entanglement of pure states is often quantified in terms of the von
Neumann
and the R\'enyi entanglement
entropies
(see, for example, Refs.\ \cite{Amico_rev2008,Calabrese_intro2009,Plenio_rev2010,Vidal2003}).
The scaling of these measures has been found very useful in distinguishing 
and
characterizing various phases of matter,
such as gapless phases and quantum critical points
described by conformal field theories
\cite{Holzhey1994,Calabrese2004,Calabrese2009},
and gapped phases such as topologically ordered phases
\cite{Kitaev_Preskill2006,Levin_Wen2006}.

On the other hand,
mixed states are described by density matrices, for example,
finite temperature states of a Hamiltonian
or states of a subsystem of a global system obtained by partial tracing.
A decent candidate to quantify the quantum entanglement within mixed states
is known as the (logarithmic) negativity of a partially transposed density matrix~\cite{Peres1996,Horodecki1996,Zyczkowski1,*Zyczkowski2,PlenioEisert1999,Vidal2002,Plenio2005}. 
 Several methods have been developed to efficiently compute the entanglement negativity in different setups ranging from   
Harmonic oscillator chains~\cite{PhysRevA.66.042327,PhysRevLett.100.080502,PhysRevA.78.012335,Anders2008,PhysRevA.77.062102,PhysRevA.80.012325} 
and quantum spin chains~\cite{PhysRevA.80.010304,PhysRevLett.105.187204,PhysRevB.81.064429,PhysRevLett.109.066403,Ruggiero_1,PhysRevA.81.032311,PhysRevA.84.062307,Mbeng2017} 
to conformal field theories (CFTs)~\cite{Calabrese2012,Calabrese2013,Ruggiero_2,Alba2018} and massive quantum field theories~\cite{Blondeau2016}.
The entanglement negativity of topologically ordered phases in (2+1) dimensions were also studied~\cite{Wen2016_1,Wen2016_2,PhysRevA.88.042319,PhysRevA.88.042318}. 
The applicability of the negativity in characterizing finite-temperature systems~\cite{Calabrese_Ft2015,Eisler_Neq,Sherman2016,Park2017}
and out-of-equilibrium scenarios~\cite{Eisler_Neq,Coser_quench2014,Hoogeveen2015, PhysRevB.92.075109} was also investigated. 
In addition to these literatures, there are other useful numerical frameworks to evaluate the entanglement negativity such as tree tensor network \cite{Clabrese_network2013}, Monte Carlo implementation of partial transpose~\cite{Alba2013,PhysRevB.90.064401}, and rational interpolations \cite{Nobili2015}.

We emphasize that the systems studied in all these works above
are made out of distinguishable objects (qubits, spins, etc.) or bosons where local operators acting on different parts of the system commute.
As it turned out,
defining an analog of entanglement negativity (and logarithmic negativity)
for fermionic systems,
and in particular partial transpose of fermionic density matrices,
is not simply a straightforward extension of what is known for bosonic systems, because of the Fermi statistics.

One possible approach to define partial transpose for fermionic systems would be to use a particular basis such as the occupation
number basis, and simply adopt the definition of the bosonic partial  transpose
-- we simply ignore any fermion sign which may arise when
we rearrange fermionic operators.
In some cases, this procedure is equivalent to use 
a Jordan-Wigner transformation (and alike) to map fermionic systems to bosonic
counterparts
and then take the bosonic partial transpose.
This was the working definition of partial transpose for fermionic systems
in early works~\cite{Eisler2015, Coser2015_1,*Coser2016_1,*Coser2016_2,PhysRevB.93.115148,PoYao2016,Herzog2016}.

It was, however, noted that if this definition of partial transpose is used to define entanglement
negativity for fermioic systems,
it fails to capture important fermionic quantum correlations. 
For example,  
with this definition of entanglement negativity, there is no entanglement
in the topological phase of the Kitaev Majorana chain~\cite{Shap_pTR}.

It was also observed that, with the above definition,
the partial transpose may turn Gaussian states (e.g., ground states or thermal
states of non-interacting fermion systems) into non-Gaussian states. 
More specifically, 
the partially transposed density matrix
would be a linear combination of two Gaussian operators,
which in general do not commute with each other.
Therefore,
the spectrum of the partially transposed density matrix
(and hence entanglement negativity)
cannot be computed easily even for noninteracting fermionic systems.
This may be regarded just as a technical difficulty,
which is in contrast with
(i) bosonic cases where the partial transpose of Gaussian states remains Gaussian~\cite{Simon2000,PhysRevA.66.042327,PhysRevA.71.032349}
and
(ii) the calculations of other entanglement measures (von-Neumann or R\'enyi
entanglement entropies) for non-interacting fermion systems where
the spectrum of Gaussian density matrices can be
easily
obtained from covariance matrices
\cite{Peschel_Eisler2009},
\footnote{
There are more issues than we just pointed out here for ``borrowing'' the
 bosonic definition of partial transpose to discuss fermionic systems.
 See also Sec.\ \ref{sec:properties}. 
}.

In Refs.~\cite{Shap_unoriented,Shap_pTR,Shiozaki_antiunitary},
we introduced a partial transpose operation that can be applied to fermionic
systems, in short, fermionic partial transpose. 
One of our motivations was, aside from quantifying entanglement,
to construct topological invariants that can diagnose and characterize
fermionic symmetry-protected topological (SPT) phases of matter protected by
 anti-unitary symmetries of various kinds.  
Examples of SPT phases studied in Refs.~\cite{Shap_unoriented,Shiozaki_antiunitary}
include time-reversal symmetric topological insulators in (2+1) dimensions (e.g., quantum spin Hall insulators),
and time-reversal symmetric topological superconductors in (1+1) dimensions 
(e.g., the Kitaev chain with time-reversal symmetry). In Ref.~\cite{Shap_pTR},
we use the fermionic partial transpose to define
a fermionic analog of (logarithmic) entanglement negativity,
which we call the fermionic entanglement negativity or simply (and loosely) the entanglement negativity when there is no confusion.


It is worth mentioning what led us to the definition of fermionic partial transpose in our earlier works.
Our construction of many-body topological invariants
follows closely the topological quantum field theory (TQFT) description of fermionic SPT phases.
In particular, to detect SPT phases protected by time-reversal symmetry (or
orientation-reversing symmetry in general), one needs to consider
path-integral on non-orientable spacetime of various kinds. 
For the case of bosonic SPT phases, it was noted that to
effectively realize non-orientable spacetime in the canonical (operator) formalism,
one can make use of partial transpose
\cite{PTurner_detect1d,Shiozaki-Ryu}. 
Now, we expect that the same strategy should work to detect fermionic SPT phases
and construct many-body topological invariants for them (see also the parallel study by Bultinck~\emph{et.~al.}~\cite{Bultinck2017}).  
TQFTs that can describe fermionic SPT phases protected by time-reversal symmetry
should involve, once again, path-integral on non-orientable spacetime. 
If there is a sensible definition of partial transpose for fermionic systems,
the many-body topological invariant constructed from it should agree with
the expected result from TQFTs on non-orientable manifolds.
This comparison with TQFTs guided us to look for a proper definition of
fermionic partial transpose~\footnote{For field theory experts, to define
the kinds of TQFTs relevant to fermionic SPT phases protected by time-reversal symmetry,
one needs to specify an appropriate Pin structure (or proper generalization
thereof).
The fermionic partial transpose should implement 
something equivalent in the canonical formalism.}.

The fact that the bosonic partial transpose fails to capture quantum correlations in
fermionic systems and that the fermionic partial transpose needs to be used instead
suggests that quantum entanglement in fermionic systems is fundamentally different from the bosonic one.
In this work, we would like to study fermionic entanglement negativity from a perspective more inclined to quantum information theory, unlike our previous studies about spacetime significance of partial transpose, TQFTs, and CFTs~\cite{Shap_unoriented,Shap_pTR,Shiozaki_antiunitary}.
To this end, we present a systematic investigation of properties of fermionic entanglement negativity among which are  the monotonicity under local quantum operations and classical communications (LOCC) (see Refs.~\cite{Vidal2002,Plenio2005,Eisert_thesis} for the monotonicity of  negativity in qubit systems), additivity, and invariance under a local unitary. 

Our paper is organized as follows: In Sec.~\ref{sec:preliminary remarks}, we review the structure of Hilbert space of local fermionic modes and provide important definitions including separable fermionic states, the fermionic partial transpose, and the fermionic entanglement negativity.
In Sec.~\ref{sec:properties},  we check that the fermionic entanglement negativity 
satisfies usual properties expected for an entanglement measure.
We provide canonical examples of entangled states for two and three fermionic modes in Sec.~\ref{sec:examples}.
In Secs.\ref{sec:class two} and \ref{sec:class three}, we use the entanglement negativity to distinguish different types of bipartite and tripartite entangled states in systems of two and three fermionic modes.
We close our discussion by final remarks in Sec.~\ref{sec:conclusions}.


\section{Preliminary remarks}
\label{sec:preliminary remarks}

We start by briefly reviewing the structure of the Hilbert space of
local fermionic modes and define the algebras of physical operators,
which include density matrix operators and any physical manipulation of Fermi systems.
Subsequently, we present
the definition of fermionic partial transpose and the corresponding entanglement
measure,
which we call entanglement negativity by analogy to the partial transpose in bosonic (qubit) systems.

Let $\mathcal{H}$ be the Hilbert space of a quantum system. The algebra ${\cal G}$ of linear bounded
 operators on the Hilbert space is generally characterized by $C^*$ algebras. Physical operators refer to a subset of linear operators whose algebra is denoted by $\textbf{L}(\Hi)$. If the Hilbert space admits superselection sectors, i.e., $\Hi= \oplus_j \Hi_j$, then the algebra of physical operators can be represented as $\oplus_j \textbf{L}(\Hi_j)$, since physical operators do not mix the superselection sectors.

{\bf Local fermionic modes:}~The case of interest in this paper is local fermionic modes where the Hilbert space is a Fock space  associated with a system of $N$ fermionic degrees of freedom (``sites") $j=1,\cdots,N$. The Fock space is spanned by $2^N$ basis vectors $\ket{n_1,\cdots,n_N}$ where $n_j=0,1$ is the occupation number of $j$th site. 
Any linear operator can be defined in terms of creation and annihilation operators $f_j^\dag$ and $f_j$   which act on the basis vectors as 
\begin{align}
  & f_j \ket{n_1,\cdots,n_{j-1},1,n_{j+1},\cdots,n_N}
    \nonumber \\
&\quad = (-1)^{\sum_{i=1}^{j-1}n_i} \ket{n_1,\cdots,n_{j-1},0,n_{j+1},\cdots,n_N}, \\
& f_j \ket{n_1,\cdots,n_{j-1},0,n_{j+1},\cdots,n_N} =0,
\end{align}
and $f_j^\dag$ is the Hermitian conjugate.
Hence, the creation and annihilation operators obey the anticommutation relations
\begin{align}
f_j^{\ } f_k^{\dag}+f_k^{\dag} f_j^{\ } &=\delta_{jk}, \\ 
f_j^{\ } f_k+f_k^{\ } f_j&=f_j^{\dag} f^{\dag}_k+f_k^{\dag} f^{\dag}_j=0
\end{align}
which generate the algebra ${\cal G}$.

We define the fermion-number parity for a basis vector by $(\sum_j n_j) \mod 2$. The fermion-number parity can also be measured by applying the operator
\begin{align} \label{eq:parity_op}
(-1)^F := (-1)^{\sum_j n_j},
\end{align}
which assigns a $\Z_2$ index to each basis vector.
In our paper, we assume a global fermion-number parity symmetry as a fundamental symmetry in  fermionic systems. This is so, simply because Hamiltonians of our interest are always bosonic operators,
and commute with the fermion number parity operator~\cite{Wick1952}.
As a result, the Hilbert space splits into two sectors $\Hi=\Hi_ 0 \oplus \Hi_1$, corresponding to the states with even (odd) number of fermions, respectively, where we can write $(-1)^F \ket{\phi}=\pm\ket{\phi}$ for $\ket{\phi} \in \Hi_0\ ( \ket{\phi} \in \Hi_1)$. In other words, the Hilbert space is $\Z_2$ graded~\cite{FidkowskiKitaev2}. 

Similarly, the operator algebra is $\Z_2$ graded as well, ${\cal G}={\cal G}_0 \oplus {\cal G}_1$. Namely, an operator $X$ is fermion-number parity even (odd) if it preserves (changes) the number of fermions modulo 2,
\begin{align}
(-1)^F X (-1)^F = (-1)^r  X, \quad \text{for} \ \ X \in {\cal G}_r \ (r=0,1).
\end{align}
This simply means that the algebra of operators ${\cal G}_0\ ({\cal G}_1)$ is spanned by products of even (odd) number of $f_j$ and $f_j^\dag$.

{\bf Physical operators:}~The physical operators are those operators which preserve the fermion-number parity symmetry~\cite{Bravyi2002,Banuls2007,Zimboras_fparity,Benatti2014}
and act on each fermion-number parity sector separately, i.e.,
\begin{align}
{\cal G}_0 = \textbf{L}(\Hi_0) \oplus \textbf{L}(\Hi_1).
\end{align}
It is evident that no physical operator can change the occupation number locally on a single site as it changes the fermion-number parity. The Hamiltonian is an example of a physical operator. 

{\bf Density matrices:}~Another important class of physical operators is density
matrix operators. The state of a system is described by a density matrix $\rho$
defined on the Hilbert space $\Hi$ and satisfies the following three conditions:
first, it is Hermitian $\rho^\dag=\rho$, second, $\rho$ is positive
semi-definite, and third, $\Tr\, \rho=1$. A system is said to be in a pure state, if $\rho^2=\rho$, otherwise, it is in a mixed state. 
Physically, mixed states represent finite-temperature states
of systems, or reduced density matrices obtained from a given pure state by taking the partial trace.
We denote the set of state density matrices by ${\cal S}(\Hi)={\cal S}(\Hi_0)\oplus {\cal S}(\Hi_1)$. 
 
{\bf Unitary operators:}~Unitary operators $U \in {\cal G}$ form a subset of linear operators which satisfy $UU^\dag=U^\dag U=\mathbb{I}$ where $\mathbb{I}$ is the identity operator on $\Hi$. In this paper, we encounter physical unitary operators $U \in {\cal G}_0$. An example of a physical unitary operator is the time-evolution under a Hamiltonian $H$, i.e., $U=\exp(-i H t)$.

{\bf Bipartite systems:}~In order to define a bipartite entanglement, we divide the system of $N$ fermion sites into  subsystems $A$ and $B$ with $m_A$ and $m_B$ sites, respectively, where $m_A+m_B=N$. 
The Hilbert space is then factorized as a tensor product,
\begin{align}
\Hi =\Hi^A \otimes \Hi^B.
\end{align}
Accordingly, we can define subalgebras of \emph{local} operators acting on each Hilbert subspace. However, these subalgebras, denoted by ${\cal G}(\Hi^i), \ i=A,B $, contain both fermion-number parity even and odd terms and do not commute with each other in general. 
Hence, in order to decompose the algebra of operators, we define the graded tensor product as 
\begin{align}
XY= X \otimes_{\text{gr}}  Y,  \quad YX=(-1)^{xy} X \otimes_{\text{gr}}  Y, 
\end{align}
where $X\equiv X  \otimes_{\text{gr}}  \mathbb{I}_B$ and $Y\equiv \mathbb{I}_A \otimes_{\text{gr}} Y$ are local operators in ${\cal G}$ and $X \in {\cal G}_{x}(\Hi^A)$ and $Y \in {\cal G}_{y}(\Hi^B)$ are two elements of the subalgebras $x,y=0$ or $1$ corresponding to even or odd sectors.
So, the algebra of linear operators admits the following decomposition
 \begin{align}
{\cal G}= {\cal G}(\Hi^A) \otimes_{\text{gr}}  {\cal G}(\Hi^B).
\end{align}
Similar to the full Hilbert space,  we may define {local} physical operators
acting on each Hilbert subspace as a set of operators which preserve the
fermion-number parity within that Hilbert subspace.
We should note that the space of physical operators on the full Hilbert space is larger than the space spanned by the tensor product of local physical operators on Hilbert subspaces, ${\cal G}_0 (\Hi^A)  \otimes  {\cal G}_0 (\Hi^B) \subseteq {\cal G}_0$. In fact, we have 
\begin{align}
{\cal G}_0 = ({\cal G}_0 (\Hi^A)  \otimes  {\cal G}_0 (\Hi^B)) \cup ({\cal G}_1 (\Hi^A) \otimes_{\text{gr}} {\cal G}_1 (\Hi^B)),
\end{align}
where we write the usual tensor product for the product of local physical observables as they always commute.
In the case of density matrices ${\cal S}(\Hi)$, this means that ${\cal S} (\Hi^A)  \otimes  {\cal S} (\Hi^B) \subseteq {\cal S}(\Hi)$, where ${\cal S} (\Hi^A)$ and ${\cal S} (\Hi^B)$ refer to the set of local density matrices. As we see below, this is essential for the existence of the space of entangled states.

{\bf Separable states:}~Any element $\rho_{\text{sep}} $ of the set ${\cal S}_{\text{sep}}={\cal S} (\Hi^A)  \otimes  {\cal S} (\Hi^B)$ can be written as 
\begin{align} \label{eq:separability}
\rho_{\text{sep}}= \sum_i w_i\ \rho_{A,i} \otimes \rho_{B,i},
\end{align}
where $w_i$ are real positive coefficients satisfying $\sum_i w_i=1$, and $\rho_{A,i}\in {\cal S} (\Hi^A)$, $\rho_{B,i} \in {\cal S} (\Hi^B)$. 
Hence, ${\cal S}_{\text{sep}}$ defines a convex set of classically correlated
states, which are also known as separable states~\cite{Peres1996,Horodecki1996,Zyczkowski1,*Zyczkowski2}.
Since these states are classically correlated, one should not expect any quantum
entanglement.
The above condition is equivalent to the separability criterion defined with reference to the bipartitioned operator algebra~\cite{Benatti2014}.

The entanglement negativity (based on partial transpose) was originally proposed as a necessary condition for a bosonic state to be separable~\cite{Peres1996,Horodecki1996}; that is, if a density matrix is separable, the negativity will vanish.
In what follows, we introduce an analog of partial transpose for fermionic density matrices. 

{\bf Majorana representation:}~For the purpose of algebraic derivation of some properties of the fermionic negativity later in the paper, it proves convenient to write the partial transpose of operators in a basis of real (Majorana) fermion operators.
For a fermionic Fock space ${\cal H}$ generated by $N$ local fermionic modes $f_j$, 
we define real fermion operators by 
\begin{align}
c_{2j-1}=f^{\dag}_j+f_j, \quad 
c_{2j}=-i(f^{\dag}_j-f_j), \quad 
j=1, \dots, N. 
\label{eq:real_fermion}
\end{align}
These operators satisfy the commutation relation
\begin{align}
c_j c_k + c_k c_j = 2 \delta_{jk}.
\end{align}
The operator algebra generated by the Majorana operators acting on the Hilbert space of $2N$ Majorana sites is known as Clifford algebra and any operator $X \in {\cal G}$ acting on this space can be expanded in terms of $c_j$, $j=1,\cdots,2N$,
\begin{align}
X
= \sum_{k=1}^{2N} \sum_{p_1<p_2 \cdots <p_k} X_{p_1 \cdots p_k} c_{p_1} \cdots c_{p_k}, 
\label{eq:op_expand}
\end{align}
where $X_{p_1 \dots p_k}$ are complex numbers and fully antisymmetric under permutations of $\{1, \dots, k\}$. When $X$ corresponds to a physical operator, $X \in {\cal G}_0$,  the expansion only contains even number of Majorana operators, i.e., $k$ is even.
For instance, a generic density matrix
$\rho \in \mathcal{S}(\Hi^A\otimes \Hi^B)$
can be expanded as
\begin{align}
\rho = \sum_{k_1,k_2}^{k_1+k_2 = {\rm even}} \rho_{p_1 \cdots p_{k_1}, q_1 \cdots q_{k_2}} a_{p_1} \cdots a_{p_{k_1}} b_{q_1} \cdots b_{q_{k_2}},
\end{align}
where
$\{ a_j \}$ and $\{ b_j \}$ are Majorana operators associated with $\mathcal{H}^A$
and $\mathcal{H}^B$, respectively, 
and the even fermion-number parity condition is shown by the condition $k_1+k_2 = {\rm even}$.

{\bf Transpose in Majorana representation:}~
A standard anti-automorphism $X \mapsto X^{\sT}$ of a Clifford algebra is defined
by reversing
the ordering of generators 
in the expansion \eqref{eq:op_expand} as in
$(c_{p_1}c_{p_2} \dots c_{p_k})^{\sf T} = c_{p_k} \cdots c_{p_2}c_{p_1}$.
Given $X,Y \in {\cal G}$, this operation is involutive $(X^{\sf T})^{\sT}=X$, linear $(zX)^{\sT}=z X^{\sT}$ for a complex number $z$,
and satisfy $(XY)^{\sT}=Y^{\sT}X^{\sT}$.
We shall call it transpose. It is easy to show that for density matrices $\rho \in {\cal S}(\Hi)$ the spectrum of $\rho$ and $\rho^{\sT}$ are identical~\cite{Shap_pTR}.

{\bf Fermionic partial transpose in Majorana representation:}~
The fermionic partial transpose is defined only in the subalgebra of physical operators ${\cal G}_0$. For $X \in {\cal G}_0$, it is given by~\cite{Shap_pTR,Shiozaki_antiunitary}
\begin{align}
 X^{\sT_A} := \sum_{k_1,k_2}^{k_1+k_2 = {\rm even}} X_{p_1 \cdots p_{k_1}, q_1 \cdots q_{k_2}} i^{k_1} a_{p_1} \cdots a_{p_{k_1}} b_{q_1} \cdots b_{q_{k_2}},
\label{eq:fermion_pt}
\end{align}
and similarly for $ X^{\sT_B}$. The partial transpose is consistent with the full transpose, since 
first, taking successive partial transpose with respect to the two subsystems is identical to taking the full transpose,~\footnote{This is a nice property of our definition of a partial transpose.  Although it is natural to expect that such identity should hold, it is not necessary. In other words, a definition of partial transpose can still be considered consistent with the full transpose as long as it satisfies $(X^{\sT_A})^{\sT_B}=U X^{\sT} U^\dag$ where $U$ is a unitary operator.}
\begin{align}
(X^{\sT_A})^{\sT_B}=X^{\sT},
\end{align}
and second, the identity operator is invariant under the partial transpose, 
\begin{align}
(\mathbb{I})^{\sf T_A} = \mathbb{I}.
\end{align}
In addition, the definition (\ref{eq:fermion_pt}) implies that
\begin{align}
(X^{\sf T_A})^{\sf T_A}=(-1)^{F_A} X (-1)^{F_A},
\end{align}
where $(-1)^{F_A}$ is the fermion-number parity operator (\ref{eq:parity_op}) defined within the Hilbert subspace $F_A=\sum_{j\in A} f_j^\dag f_j$. This property reflects the fact that the fermionic partial transpose is related to the action of time-reversal operator of spinless fermions in the Euclidean spacetime~\cite{Shap_unoriented}.

{\bf Fermionic partial transpose in occupation number basis:}~ The difference between bosonic partial transpose (that is a matrix partial transposition) and the fermionic one is most obvious in the occupation-number basis. Due to anti-commutation of local fermionic modes, let us fix our convention by defining the ``normal-ordered" occupation-number basis as
\begin{align}
&\ket{\{ n_j \}_{j \in A} , \{n_j\}_{j \in B}} \nonumber \\
=& (f_{j_1}^{\dag})^{n_{j_1}} \cdots (f_{j_{m_A}}^{\dag})^{n_{j_{m_A}}} \cdots (f_{j_{m_B}'}^{\dag})^{n_{j_{m_B}'}} \ket{0}
\end{align}
where $n_j$'s are occupation numbers for the subsystems $A$ and $B$,
respectively, and
$\ket{0}$ 
denotes the vacuum state where all $n_j$'s are zero.
 Normal-ordering in this representation simply means that all fermionic degrees of freedom within each subsystem are clustered together. Such normal-ordering can always be achieved for any given set of local fermionic sites, after shuffling around the fermion operators and keeping track of minus signs due to the anti-commutation relation. For instance, consider four fermion sites living on a chain and label them from $1$ to $4$. Let us partition it such that sites $1$ and $4$ belong to the subsystem $A$ and sites $2$ and $3$ belong to the subsystem $B$. To normal-order a state like $\ket{1011}=f_1^\dag f_3^\dag f_4^\dag \ket{0}$ means to bring the fermion site $4$ next to site $1$ and write it in this form $\ket{\{ n_j \}_{j \in A} , \{n_j\}_{j \in B}}$, i.e., $\ket{11,01}=f_1^\dag f_4^\dag f_3^\dag \ket{0}=-f_1^\dag f_3^\dag f_4^\dag \ket{0}$ which differs from the original (spatial) ordering by a minus sign.

Hence, a normal-ordered density matrix $\rho$ in this basis reads as
\begin{align}
\sum_{ n_j  ,  \bar n_j}^{ \sum_j n_j +\bar n_j=\text{even}}  \rho (\{ n_j \} , \{\bar n_j\}) \ket{\{ n_j \}_{A} , \{n_j\}_{B}} \bra{\{ \bar n_j \}_{A}, \{ \bar n_j \}_{B} } 
\end{align}
where $\rho(\{ n_j \} , \{\bar n_j\})=\bra{\{  n_j \}_{A}, \{  n_j \}_{B} } \rho \ket{\{ \bar n_j \}_{A} , \{\bar n_j\}_{B}}$ and the constraint $\sum_j n_j +\bar n_j=\text{even}$ implies $\rho \in {\cal S}(\Hi^A\otimes \Hi^B)$.
The fermionic partial transpose (\ref{eq:fermion_pt}) of a basis vector is then given by 
\begin{align}
&\left( \ket{\{ n_j \}_{A} , \{n_j\}_{B}} \bra{\{ \bar n_j \}_{A}, \{ \bar n_j \}_{B} } \right)^{\sT_A}
\nonumber \\
=&(-1)^{\phi(\{n_j\}, \{\bar n_j\})} U_{A}^\dag \ket{\{ \bar n_j \}_{A} , \{n_j\}_{B}} \bra{\{ n_j \}_{A}, \{ \bar n_j \}_{B} } U_{A},
\label{eq:app_f_21}
\end{align}
where the phase factor is
\begin{align}
\phi(\{n_j\}, \{\bar n_j\}) =& \frac{[(\tau_A+\bar{\tau}_A)\ \text{mod}\ 2]}{2} + (\tau_A+\bar{\tau}_A)(\tau_B+\bar{\tau}_B)
\end{align}
in which
$\tau_{A(B)}=\sum_{j\in A(B)} n_j$  and $\bar{\tau}_{A(B)}=\sum_{j\in A (B)} \bar{n}_j$, are the number of occupied states in each subsystem. Note that $U_{A}$ is a unitary transformation (partial particle-hole transformation) $U_A=\prod_{j \in A} c_{2j-1}$ acting on $\Hi^A$.  As far as the entanglement negativity is concerned $U_{A}^\dag \rho^{\sf T_A} U_A$ and $\rho^{\sf T_A}$ have identical eigenvalues which lead to the same value for the entanglement negativity (defined below). 

It is now evident that the fermionic definition (\ref{eq:app_f_21}) is distinct from the bosonic partial transpose~\cite{Eisler2015} (which is just exchanging the states of subsystem $A$) due to presence of a phase factor.

{\bf Fermionic entanglement negativity:}
For a pure state $\ket{\Psi} \in \Hi^A\otimes \Hi^B$, the bipartite entanglement is computed by the von Neumann entanglement entropy
\begin{align}
\label{eq:vN}
S_{\text{vN}}(\rho_A)= -\Tr (\rho_A \log \rho_A),
\end{align}
and the R\'enyi entropies
\begin{align}\label{eq:Renyi}
S_n(\rho_A)= \frac{1}{1-n} \log \Tr (\rho_A^n),
\end{align}
 in terms of reduced density matrix $\rho_A=\Tr_B(\ket{\Psi}\bra{\Psi})$. Notice that $S(\rho_A)=S(\rho_B)$ and $S_n(\rho_A)=S_n(\rho_B)$.
 
For a mixed state $\rho \in {\cal S}(\Hi^A\otimes \Hi^B)$,  entanglement could be quantified by the difference between the  von Neumann or R\'enyi entropies of different partitions. This way, one can define the mutual information,
\begin{align}
I_{\text{vN}}(\rho) &= S_{\text{vN}}(\rho_A)+ S_{\text{vN}}(\rho_B)- S_{\text{vN}}(\rho), \\
I_n(\rho) &= S_n(\rho_A)+ S_n(\rho_B)- S_n(\rho), 
\end{align}
where $\rho_A= \Tr_B\rho$ and $\rho_B= \Tr_A\rho$. However, the mutual information is known to count  classical correlations and is not an entanglement measure for generic mixed states. 

A general recipe to quantify entanglement in mixed states is through the convex roof extension of pure state entanglement measures~(\ref{eq:vN}).~A mixed state density matrix is decomposed in terms of a set of pure states as $\rho=\sum_i p_i \ket{\psi_i}\bra{\psi_i}$  with probabilities $p_i$ and entanglement of formation~\cite{Bennett_formation} is defined as
\begin{align}
E_F(\rho)=\inf_{\{p_i,\psi_i \}} \sum_i p_i E(\rho_i),
\end{align}
where $\rho_i=\Tr_B(\ket{\psi_i}\bra{\psi_i})$ and $E(\rho_i)$ is the von Neumann entanglement entropy of the reduced density matrix or any measure of pure-state entanglement that is monotonically related to $S_{\text{vN}}(\rho_i)$. In practice, constructing all possible decompositions of a general mixed state, i.e.,~$\{p_i,\psi_i \}$, and finding the infimum of this ensemble are rather arduous for systems larger than a two-qubits system~\cite{Wootters1}. In this regard, $E_F(\rho)$ is not a computationally desirable way to quantify the entanglement in mixed states.

Another entanglement quantity which can be rather easily computed for mixed states is the entanglement negativity associated with the partial transpose. In an earlier paper~\cite{Shap_pTR}, we introduce an analog of the negativity for fermions
\begin{align} \label{eq:neg}
{\cal N}(\rho)= \frac{\norm{\rho^{\sT_A}}-1}{2},
\end{align}
as well as an analog of the logarithmic negativity by
\begin{align} \label{eq:logneg}
{\cal E}(\rho)= \log \norm{\rho^{\sT_A}},
\end{align}
where  
\begin{align}  \label{eq:trnorm}
\norm{A}= \Tr \sqrt{A A^\dag}.
\end{align}
is the trace norm.
 Note that for a pure state $\rho=\ket{\Psi}\bra{\Psi}$,  ${\cal E}(\rho)=S_{1/2}(\rho_A)$ where $S_{1/2}(\rho_A)$ is the  $1/2$-R\'enyi entropy defined in (\ref{eq:Renyi}) for $n=1/2$.
Higher moments of fermionic partial transpose can be defined as
\begin{align} \label{eq:pathTR}
{\cal E}_n :=\left\{ \begin{array}{ll}
\log \Tr (\rho^{\sT_A} \rho^{\sT_A\dag}  \cdots \rho^{\sT_A}\rho^{\sT_A\dag})& \ \ n\  \text{even}, \\
\log \Tr (\rho^{\sT_A} \rho^{\sT_A\dag} \cdots \rho^{\sT_A}) &  \ \ n\  \text{odd}
\end{array} \right.
\end{align}
which are also noted as the upper bound for the bosonic negativity~\cite{Herzog2016,Eisert2016}.

We close this section by remarking that the presence of phase factor in the fermionic partial transpose (\ref{eq:app_f_21}) is quite fundamental and becomes more apparent in the path-integral formalism of moments of partial transpose (\ref{eq:pathTR}). The fermionic partial transpose is equivalent to the partition function on spacetime manifold with a fixed spin structure whereas the bosonic partial transpose is equivalent to a sum over all possible spin structures (see Ref.~\cite{Shap_pTR} for an extended discussion on this.)



\section{General properties of fermionic entanglement negativity}
\label{sec:properties}

In order for an entanglement measure to be useful, it should satisfy several requirements~\cite{Vedral1997,Brus2002,Eisert_thesis}. In this section,  we investigate how the fermionic entanglement negativity treats each requirement.


Before we discuss the properties, we investigate some essential identities
for partially transposed fermionic operators which will be used later in our derivations.
These identities may seem rather trivial for bosonic partial transpose where the partial
transpose is solely a matrix partial transposition. However, for fermionic
partial transpose we multiply by some complex phase factor as we take the partial
transpose, and it is not fully clear that the above identities may hold.
As we show below, the fermionic partial transpose does satisfy all of the
expected identities for a set of states ${\cal S}(\Hi^A\otimes \Hi^B)$ and physical operators in the algebra ${\cal G}_0$ which preserve the fermion-number parity.


\begin{prop}
\label{prop:ABrT_1}
For local physical operators $X_s \in {\cal G}_0(\mathcal{H}^s)\ s=A,B$ which act only on the $A$ and $B$ subsystems  and a state density matrix $\rho  \in {\cal S}(\Hi^A \otimes \Hi^B)$, we have
\begin{align}
\label{eq:ABrT_1}
[\rho (X_A\otimes X_B)]^{\sT_A}= (X_A^{\sT}\otimes \mathbb{I}_B) \rho^{\sT_A} (\mathbb{I}_A \otimes X_B) , 
\end{align}
or
\begin{align}
\label{eq:ABrT_2}
[(X_A\otimes X_B) \rho]^{\sT_A}=  (\mathbb{I}_A\otimes X_B) \rho^{\sT_A} (X_A^{\sT}\otimes \mathbb{I}_B) , 
\end{align}
where $\mathbb{I}_A$ and $\mathbb{I}_B$ are the identity operators on $\mathcal{H}^A$ and $\mathcal{H}^B$, respectively.
\end{prop}

A proof of this proposition is given in Appendix~\ref{sec:useful identities}. 

\begin{prop}
For local physical operators $X_s,Y_s \in {\cal G}_0(\mathcal{H}^s),\ s=A,B$ and a state density matrix $\rho  \in {\cal S}(\mathcal{H}^A \otimes \mathcal{H}^B)$, we have
\begin{align}
\label{eq:ABCDrT_1}
[ (X_A\otimes X_B) \rho (Y_A\otimes Y_B)]^{\sT_A} &= (Y_A^{\sT}\otimes X_B) \rho^{\sT_A} (X_A^{\sT}\otimes Y_B), \\
\label{eq:ABCDrT_2}
[ (X_A\otimes X_B) \rho (Y_A \otimes Y_B)]^{\sT_B} &= (X_A \otimes Y_B^{\sT}) \rho^{\sT_B} (Y_A \otimes X_B^{\sT}).
\end{align}
\end{prop}

\noindent Proof:

Below, we show (\ref{eq:ABCDrT_1}). Equation (\ref{eq:ABCDrT_2}) can be derived similarly.
\begin{align}
&( (X_A\otimes X_B) [\rho (Y_A \otimes Y_B)])^{\sT_A} \nonumber \\
&=  (\mathbb{I}_A \otimes X_B) [\rho (Y_A\otimes Y_B)]^{\sT_A} (X_A^{\sT} \otimes \mathbb{I}_B)
\\
&= (\mathbb{I}_A \otimes B) (Y_A^{\sT} \otimes \mathbb{I}_B) \rho^{\sT_A} (\mathbb{I}_A\otimes Y_B) (X_A^{\sT} \otimes \mathbb{I}_B) \\
&=  (Y_A^{\sT} \otimes X_B) \rho^{\sT_{A}} (X_A^{\sT} \otimes Y_B) 
\end{align}
where in the first and second lines we use (\ref{eq:ABrT_2}) and (\ref{eq:ABrT_1}), respectively.
\hfill $\blacksquare$


Now, we study the standard properties of an ideal entanglement measure one by one for the fermionic entanglement negativity. 

\subsection{Zero entanglement for separable states}
\label{sec:separability}

The concept of entanglement negativity is intimately related to the separability criterion of density matrices and various studies show the versatility of this criterion for detecting quantum entanglement~\cite{Simon2000,PhysRevLett.86.3658,PhysRevLett.87.167904,RevModPhys.84.621,Banuls2007,Banuls2009,Botero200439,Benatti2012a,Benatti2014}.
Before we delve into details of separability in fermionic systems, let us make a few remarks about the relation between negativity and separability criterion in bosonic systems.
 Despite the equivalency of the entanglement negativity and separability criterion for a two-qubit system~\cite{Peres1996,Horodecki1996} and Gaussian states of two-mode bosonic system~\cite{Simon2000,PhysRevLett.86.3658}, it has been established that in general for larger (bosonic) systems, vanishing negativity is a necessary (but not sufficient) condition for separability. In other words, there exist cases~\cite{Horodecki1997} in which the state is inseparable but the eigenvalues of the partial transpose are positive or equivalently, the negativity is zero. These types of states are said to have a bound entanglement~\cite{Horodecki1998}, i.e.~their entanglement cannot be distilled~\cite{Bennett1996a,Bennett1996b,Horodecki1997_distillation} as a resource for carrying out quantum computation processes such as teleportation. We should note that these states cannot be prepared by means of LOCC on separable states. 
These observations have led to an important discovery that the subspace of states with positive partial transpose (PPT), i.e., zero negativity, is not only limited to the separable states and it contains bound entangled states as well.
Accordingly, the group of LOCC operations is enlarged to the so-called PPT operations which can be used to construct all PPT states, separable or bound entangled.
 
A crucial assumption in the original definition of the separability criterion is the tensor product structure of local Hilbert spaces, e.g.~the Hilbert space of a composite system $\Hi$ can be decomposed as $\Hi^A\otimes \Hi^B$ associated with two subsystems $A$ and $B$. 
This requirement was relaxed in a more general definition of separability in terms of a bipartition of the operator algebra \cite{Barnum2004,Barnum2005}. The operator-based separability criterion is simplified to the original separability criterion whenever locally factorizable Hilbert spaces are available.
Furthermore, the operator-based definition has been applied to systems of indistinguishable particles~\cite{Banuls2007,Banuls2009,Botero200439,Benatti2012a,Benatti2012b,Benatti2014} where the analog of an entanglement between distinguishable qubits would become an entanglement between the modes occupied by indistinguishable particles (fermions or bosons). See also Refs.~\cite{Bruschi,Kraus2018} for a discussion of the fermionic-mode entanglement in generic density matrices and Gaussian states, respectively. In the case of a particle-number conserving system of bosons, it was found that vanishing negativity and separability are equivalent for a bipartition of $N\geq 2$ bosonic modes into a one mode and the rest~\cite{Benatti2012a,Benatti2012b}.

A separable state $\rho_{\text{sep}} $ in a Fock space of local fermionic modes is defined in (\ref{eq:separability}). 
Applying the partial transpose yields
\begin{align}\label{eq:separable}
\rho_{\text{sep}}^{\sT_A}= \sum_i w_i\ \rho_{A,i}^{\sT}  \otimes \rho_{B,i},
\end{align}
where partial transpose acts as a full transpose on the local density matrix in the right-hand side.
As mentioned earlier, the full transpose does not change the spectrum of a density matrix. Hence, $\rho_{A,i}^{\sT}\in {\cal S}(\Hi^A)$, since it is a non-negative matrix with unit trace. This implies that $\rho_{\text{sep}}^{\sT_A} \in {\cal S}(\Hi^A)\otimes {\cal S}(\Hi^B)$ is a density matrix, as it is a convex combination of product states $\rho_{A,i}^{\sT}  \otimes \rho_{B,i}$. Therefore, we may write
 $\Tr |\rho_{\text{sep}}^{\sT_A}|=\Tr \rho_{\text{sep}}^{\sT_A}=1$ or ${\cal N} (\rho_{\text{sep}})=0$. 

We should note that the separability condition of fermions (\ref{eq:separability}), compared with that of bosons (qubits), has an additional constraint on local density matrices $\rho_{A,i}$ and $\rho_{B,i}$, that they are fermion-number parity even within their own Hilbert spaces. 
This property can also be derived from the separability criterion based on bipartition of operator algebras~\cite{Benatti2014}.
Remarkably, the definition of fermionic partial transpose (\ref{eq:fermion_pt}) (and full transpose) is sensitive to this property, since the full transpose only gives the same spectrum for Hermitian physical (even fermion number parity) operators. 

It is important to note that similar to the bosonic partial transpose the set of states with zero negativity is convex for the fermionic partial transpose.
\begin{thm}
The states of vanishing negativity ${\cal N}=0$ $($or equivlanetly, ${\cal E}=0$$)$ form a convex set.
\label{thm:PPT_convex}
\end{thm}

\noindent Proof:

Consider a linear interpolation between two states $\rho_1$ and $\rho_2$ with a zero negativity, i.e., $\norm{\rho_i^{\sT_A}} =1$,
\begin{align}
\rho_p= p \rho_{1} + (1-p)\rho_2,
\end{align}
where $0\leq p\leq 1$. We can then write
\begin{align}
\norm{\rho_p^{\sT_A}} =& \norm{p \rho_{1}^{\sT_A} + (1-p)\rho_2^{\sT_A}} \\
\leq & p\norm{\rho_{1}^{\sT_A}} +   (1-p) \norm{\rho_2^{\sT_A}} =1
\end{align}
where we use the triangle inequality of the norm. The only solution to the above inequality is when  $\norm{\rho_p^{\sT_A}}=1$, since $\norm{\rho_p^{\sT_A}}$ cannot be smaller than one. This means that ${\cal N}(\rho_p)=0$ for all $0\leq p\leq 1$, which is the statement of convexity.

 \hfill $\blacksquare$

Moreover, we have not encountered any example of bound entangled states in fermionic systems. We believe that this is partly due to the fermion-number parity constraint on density matrices (see also the discussion below Eq.~(\ref{eq:conj1}) in Sec.~\ref{sec:conclusions}). 
In low-dimensional Hilbert spaces, we provide theorems that forbid the existence of such states later in this paper. For example, for a system of two fermionic modes (Theorem \ref{thm:2f_sep}) or a system of one fermionic mode coupled to arbitrary fermionic Fock space (Theorem \ref{thm:bisep}), the fermionic entanglement negativity is a necessary and sufficient condition for the separability, i.e., there is no inseparable state with zero negativity.

It is important to note that if we treat a density matrix of fermions in the occupation-number basis as a bosonic entity, we may be led to wrong results.
For instance, it is possible that an inseparable density matrix of fermions expressed as a separable state within the bosonic formalism. This is, however, wrong, since the local bosonic density matrices are not legitimate fermionic density matrices (see examples in Sec.~\ref{sec:examples}). 

 \subsection{Invariance under local unitary transformation}

 Applying local unitary physical operator must not change the entanglement measure.
 A local unitary physical operation is represented by 
 \begin{align}
 \rho \to (U_A\otimes U_B) \rho (U_A^\dag \otimes  U_B^\dag),
 \end{align}
where $U_s \in {\cal G}_0({\cal H}^s), \ s=A,B$ are unitary physical operators acting on subsystems $A$ and $B$, respectively. We are to show that
\begin{align} \label{eq:mono2}
{\cal N} \left( (U_A\otimes U_B) \rho (U_A^\dag \otimes  U_B^\dag)\right) = {\cal N}(\rho),
\end{align}
or equivalently,
\begin{align} \label{eq:mono2_logN}
{\cal E} \left( (U_A\otimes U_B) \rho (U_A^\dag \otimes  U_B^\dag)\right) = {\cal E}(\rho).
\end{align}

\noindent Proof:

Using (\ref{eq:ABCDrT_1}), we write
 \begin{align}
& \norm{ \left[ (U_A\otimes U_B) \rho (U_A^\dag \otimes  U_B^\dag) \right]^{\sT_A}} \nonumber \\
 =&  \norm{  ((U_A^\dag)^{\sT} \otimes U_B) \rho^{\sT_A} ( U_A^{\sT}\otimes  U_B^\dag) } \nonumber \\
=&  \norm{  (\tilde{U}_A \otimes U_B) \rho^{\sT_A} ( \tilde{U}_A^\dag \otimes  U_B^\dag) } \nonumber \\
  =& \norm{\rho^{\sT_A}}
  \label{eq:mono2_norm}
 \end{align}
where in the second line we observe that $\tilde{U}_A=(U_A^\dag)^{\sT}$ is also a unitary physical operator, since 
\begin{align}
\tilde{U}_A \tilde{U}_A^\dag= (U_A^\dag)^{\sT} U_A^{\sT}=(U_A U_A^\dag)^{\sT}=\mathbb{I}_A.
\end{align}
This makes the overall operator $\tilde{U}_A \otimes U_B$ unitary. Lastly, in the third line of the proof we use the fact that trace norm does not change under unitary transformations. This then immediately implies (\ref{eq:mono2}) and (\ref{eq:mono2_logN}). 
 \hfill $\blacksquare$

A special case of (\ref{eq:mono2_norm}) is when we apply uni-local unitary physical operator $U_A\otimes \mathbb{I}_B$, i.e., when $U_B=\mathbb{I}_B$,
\begin{align} \label{eq:mono2_unilocal}
\norm{ (U_A\otimes \mathbb{I}_B) \rho (U_A^\dag \otimes  \mathbb{I}_B)} = \norm{\rho}.
\end{align}
 
  \subsection{Additivity}
  
  Entanglement of a composite system is equal to the sum of the entanglements of the constituting systems.
 To be more specific, let us consider two sets of local fermionic modes corresponding to the Fock spaces $\Hi$ and $\Hi'$. The combined density matrix $\rho\in {\cal S}(\Hi)\otimes  {\cal S}(\Hi')$ can be written as
 \begin{align}
 \rho=\rho_{AB} \otimes \rho'_{AB},
 \end{align}
 where $\rho_{AB} \in {\cal S}(\Hi^A\otimes\Hi^B)$ and $\rho'_{AB} \in {\cal S}(\Hi'^A\otimes\Hi'^B)$ describe entangled states in $\Hi$ and $\Hi'$, respectively.
  This situation, for example, is realized by stacking two chains where $\Hi^A$ and $\Hi'^A$ spaces belong to subsystem $A$ and $\Hi^B$ and $\Hi'^B$ spaces belong to subsystem $B$. The additivity condition requires that
  \begin{align} \label{eq:additivity}
  {\cal E} (\rho_{AB} \otimes \rho'_{AB}) = {\cal E} (\rho_{AB}) + {\cal E}(\rho'_{AB}) .
  \end{align}
  
  \noindent Proof:
  
An important property of the fermionic partial transpose (\ref{eq:fermion_pt}) is that it preserves the tensor product of fermionic density matrices and we can write
\begin{align} \label{eq:tensor_prod}
(\rho_{AB} \otimes \rho_{AB}' )^{\sT_A}=\rho_{AB}^{\sT_A} \otimes \rho_{AB}'^{\sT_A}.
\end{align}
This can be easily seen in the Majorana representation of density matrices. The additivity condition (\ref{eq:additivity}) holds immediately. A slightly different statement of additivity is that if we consider $n$ copies of a density matrix $\rho \in {\cal S}(\Hi^A\otimes \Hi^B)$, we would have
\begin{align}
{\cal E}(\rho^{\otimes n})=  n {\cal E}(\rho).
\end{align}
 \hfill $\blacksquare$

It is worth noting that the bosonic partial transpose of fermionic density matrices~\cite{Eisler2015} does not preserve the tensor product structure and hence does not satisfy the additivity condition (see Appendix~\ref{sec:bosonic_transpose}). This violation is rooted in the fact that the fermionic nature of density matrices was completely ignored in this formalism. Additivity is sometimes considered as being too strong. Instead, one defines subadditivity as
  \begin{align} \label{eq:subadditive}
 {\cal E}(\rho_{AB} \otimes \rho_{AB}') \leq {\cal E}(\rho_{AB})+  {\cal E}(\rho_{AB}').
 \end{align}
The bosonic partial transpose of fermionic density matrices does not satisfy this condition either (see example of stacking two Majorana chains in our earlier work~\cite{Shap_pTR}).
 

 \subsection{No increase (monotonicity) under LOCC}
Here, we study how the entanglement negativity behaves under the action of
local quantum operations and classical communication (LOCC) and show that it is an entanglement monotone. 

For notational clarity, we use $\rho_{AB} \in {\cal S}(\Hi^A\otimes \Hi^B)$ to denote an entangled state density matrix between subsystems $A$ and $B$ as we need to introduce a state $\rho_R$ associated with an ancilla . 
 
\subsubsection{Appending ancilla}

Appending an unentangled local ancilla $R$ must not change the entanglement measure.

This is modeled by the following process
\begin{align} 
\rho_{AB} \to ( \rho_{AB} \otimes \rho_R),
\end{align}
where we add an ancilla in an arbitrary mixed state, denoted by $\rho_R \in {\cal S}(\Hi^R)$, to our original system $\rho_{AB}$ and by local ancilla, we mean that the new global system $R \cup (A B)$ is partitioned to  $\tilde{A}=AR$ and $B$. We need to show
\begin{align} \label{eq:mono1}
{\cal N}(\rho_{AB} \otimes \rho_R)= {\cal N}(\rho_{AB}),
\end{align}
which also implies
\begin{align} \label{eq:mono1_logN}
{\cal E}(\rho_{AB} \otimes \rho_R)= {\cal E}(\rho_{AB}).
\end{align}

\noindent Proof:

Because of the new partitioning scheme and hence the corresponding enlarged Hilbert space $\Hi^R\otimes \Hi^A$ of subsystem $\tilde{A}$, now we must take the partial transpose with respect to $AR$, denoted by $(.)^{{\sT}_{\tilde A}}$. The partial transpose of the combined density matrix is then given by
\begin{align} 
  ( \rho_{AB} \otimes \rho_R)^{{\sT}_{\tilde{\sA}}}
  &= \rho_{AB}^{\sT_{\tilde{\sA}}} \otimes \rho_R^{\sT_{\tilde{\sA}}} 
\nonumber \\
&=\rho_{AB}^{\sT_A} \otimes \rho_R^{\sT}
\label{eq:anc_T}
\end{align}
where in the first identity we use the tensor product property (\ref{eq:tensor_prod}). In the second line, we write 
$ \rho_{AB}^{\sT_{\tilde{\sA}}} = \rho_{AB}^{\sT_{A}}$ since $\rho_{AB}$ only contains $A$ part of $\tilde{A}$ and $ \rho_{R}^{\sT_{\tilde{\sA}}} = \rho_{R}^{\sT}$ since $\rho_R$ is fully embedded in $\tilde{A}$. We know that a fully transposed density matrix is also a legitimate density matrix. Hence, all the eigenvalues of $\rho_R^{\sT}$ are real positive and $\norm{\rho_R^{\sT}}=\Tr(\rho_R^{\sT})=1$. Consequently, we can write
\begin{align}
 \norm{( \rho_{AB} \otimes \rho_R)^{{\sT}_{\tilde{\sA}}}} 
&=\norm{\rho_{AB}^{\sT_A} \otimes \rho_R^{\sT}} \nonumber \\
&=\norm{\rho_{AB}^{\sT_A}}\cdot \norm{\rho_R^{\sT}} \nonumber \\
&=\norm{\rho_{AB}^{\sT_A}}
  \label{eq:mono1_norm}
\end{align}
which results in (\ref{eq:mono1}).
\hfill $\blacksquare$

\subsubsection{Local projectors}

 Application of local projective measurements does not increase the entanglement measure. A local projection operator $P_s, \ s=A,B$ acting on the Hilbert space ${\Hi}^s$ is a physical operator $P \in {\cal G}_0(\Hi^s)$ which satisfies $P^2_s=P_s$.

 Consider two sets of orthogonal local projectors $\{P^\mu_A\}$ and $\{P^\mu_B\}$ on subsystems $A$ and $B$. The locally projected density matrices are
\begin{align}
\rho_{AB}(\mu)=\frac{1}{r_\mu} (P^\mu_A \otimes P^\mu_B) \rho_{AB} (P^\mu_A\otimes P^\mu_B),
\end{align}
where $r_\mu=\Tr[ (P^\mu_A \otimes P^\mu_B) \rho_{AB} (P^\mu_A\otimes P^\mu_B)]$. The completeness of the set of projectors implies 
\begin{align}
 \Tr\left[ \sum_\mu  (P^\mu_A \otimes P^\mu_B) \rho_{AB} (P^\mu_A\otimes P^\mu_B) \right]=1,
\end{align}
which means
\begin{align}
\sum_\mu r_\mu=1.
\end{align}
 The monotonicity condition can then be written as
\begin{align} \label{eq:mono3}
{\cal N}(\rho_{AB}) \geq \sum_\mu  r_\mu  {\cal N}\left(\rho_{AB}(\mu)  \right).
\end{align}
which also implies
\begin{align} \label{eq:mono3_logN}
{\cal E}(\rho_{AB}) \geq \sum_\mu  r_\mu  {\cal E}\left(\rho_{AB}(\mu)  \right),
\end{align}
using the relation ${\cal E}=\log (2{\cal N}+1)$ and the fact that $\log$ is concave.

\noindent Proof:

We begin by noting that for any set of orthogonal projectors, we have $\norm{\sum_i P_i A P_i}\leq \norm{A}$ which is a property of any unitarily invariant norm.
Hence, we have
\begin{align}
\norm{\rho_{AB}^{\sT_A}} \geq & \norm{ \sum_\mu (({P}^\mu_A)^{\sT} \otimes P^\mu_B) \rho_{AB}^{\sT_A} (
({P}^\mu_A)^{\sT} \otimes P^\mu_B) } \nonumber \\
=&   \sum_\mu \norm{ (({P}^\mu_A)^{\sT} \otimes P^\mu_B) \rho_{AB}^{\sT_A} ((P^\mu_A)^{\sT} \otimes P^\mu_B) } \nonumber \\
=&  \sum_\mu  \norm{ \left[(P^\mu_A \otimes P^\mu_B) \rho_{AB} (P^\mu_A\otimes P^\mu_B) \right]^{\sT_A} } \nonumber \\
=& \sum_\mu r_\mu \norm{\rho_{AB}(\mu)^{\sT_A}},
  \label{eq:mono3_norm}
\end{align}
where we note that both $({P}^\mu_A)^{\sT}$ and ${P}^\mu_A$ are projection operators, since $({P}^\mu_A)^2={P}^\mu_A$ and hence, $(({P}^\mu_A)^{\sT})^2=({P}^\mu_A)^{\sT}$.
Moreover, in the third line we use (\ref{eq:ABCDrT_1}) that the partial transpose acts as a full transpose on the local projectors of the first subsystem $P^\mu_A$.
\hfill $\blacksquare$

\subsubsection{Tracing out ancilla}

Tracing out an entangled ancilla must not increase the entanglement measure.

Let us consider an ancilla which is entangled to the subsystem $A$. The process of entangling an ancilla is as follows:  We put ancilla together with the subsystem $A$ which is described by $\rho_{AB}\otimes \rho_R$ where $\rho_R$ is the initial state of the ancilla. We let the subsystem $\tilde{A}=AR$ evolve under some unitary evolution $U_{AR} \in {\cal G}_0(\Hi^A\otimes \Hi^R)$. As a result, we get
\begin{align}
\rho_{\tilde{A} B}=(U_{AR}\otimes \mathbb{I}_B) \rho_{AB}\otimes \rho_R (U_{AR}^\dag\otimes \mathbb{I}_B)
\end{align}
Now, measuring the ancilla and finding it to be in a state $\ket{\mu}_R$ leads to a projected density matrix $\rho_{AB}(\mu)$,
\begin{align}
\rho_{AB}(\mu)&= \frac{1}{r_\mu} \tensor[_R]{\bra{\mu}}{} \rho_{\tilde{A} B} \ket{\mu}_R \\
&= \frac{1}{r_\mu} \tensor[_R]{\bra{\mu}}{} (U_{AR}\otimes \mathbb{I}_B) \rho_{AB}\otimes \rho_R
(U_{AR}^\dag \otimes \mathbb{I}_B) \ket{\mu}_R
\end{align}
where $r_\mu=\Tr_{AB}[ \tensor[_R]{\bra{\mu}}{} U_{AR} (\rho_{AB}\otimes \rho_R) U_{AR}^\dag \ket{\mu}_R]$ is the probability of observing the state $\rho_{AB}(\mu)$. Tracing out ancilla yields
\begin{align}
& \Tr_R\left(\rho_{\tilde{A} B}\right) \nonumber \\
=& \sum_\mu \tensor[_R]{\bra{\mu}}{} (U_{AR}\otimes \mathbb{I}_B) \rho_{AB}\otimes \rho_R (U_{AR}^\dag \otimes \mathbb{I}_B) \ket{\mu}_R \nonumber \\
=& \sum_\mu r_\mu \rho_{AB}(\mu).
\end{align}
The statement of monotonicity is that the process of entangling ancilla to subsystem $A$ and measuring the ancilla without subselection should not increase the entanglement, i.e., entanglement measure is required to satisfy
\begin{align}  \label{eq:mono4_1}
{\cal N}(\rho_{AB}) &\geq \sum_\mu  r_\mu {\cal N}( \rho_{AB}(\mu)),\\
 \label{eq:mono4_2}
 {\cal N}(\rho_{AB}) & \geq {\cal N}(\sum_\mu r_\mu \rho_{AB}(\mu)).
\end{align}
We should note that we only have the first inequality for the logarithmic negativity ${\cal E}(\rho_{AB})$  since 
${\cal E}(\rho_{AB}) $  is not a convex function. 
\begin{align}  \label{eq:mono4_logN}
{\cal E}(\rho_{AB}) &\geq \sum_\mu  r_\mu {\cal E}( \rho_{AB}(\mu)).
\end{align}

\noindent Proof:

We first note that 
\begin{align}
\norm{\rho_{\tilde{A} B}^{\sT_{\tilde{\sA}}}} &= \norm{\left((U_{AR}\otimes \mathbb{I}_B) \rho_{AB}\otimes \rho_R (U_{AR}^\dag \otimes \mathbb{I}_B) \right)^{\sT_{\tilde{\sA}}} } \nonumber \\
&=\norm{(\rho_{AB}\otimes \rho_R )^{\sT_{\tilde A}} } \nonumber \\
&= \norm{\rho_{AB}^{\sT_{A}}} 
\label{eq:mono3_proof_1}
\end{align}
where the second and third identities follow from the application of uni-local unitary (\ref{eq:mono2_unilocal}) and the first type of LOCC (\ref{eq:mono1_norm}), respectively. On the other hand, we may write
\begin{align}
 \norm{\rho_{AB}^{\sT_{A}}}  = \norm{\rho_{\tilde{A} B}^{\sT_{\tilde{\sA}}}}  
 &\geq  \sum_\mu \norm{[(P^\mu_{R}\otimes \mathbb{I}_{AB}) \rho_{\tilde{A} B  }(P^\mu_{R}\otimes \mathbb{I}_{AB})]^{\sT_{\tilde{\sA}}}  } \nonumber \\
 &= \sum_\mu \norm{ [ \ket{\mu}_R \tensor[_R]{\bra{\mu}}{} \rho_{\tilde{A} B} \ket{\mu}_R  \tensor[_R]{\bra{\mu}}{} ]^{\sT_{\tilde{\sA}}}  } 
 \nonumber \\
 &= \sum_\mu r_\mu \norm{ [\rho_{AB}(\mu) \otimes \ket{\mu}_R \tensor[_R]{\bra{\mu}}{} ]^{\sT_{\tilde{\sA}}}  } \nonumber \\
&= \sum_\mu r_\mu \norm{[\rho_{AB}(\mu)]^{\sT_A}}
\label{eq:mono3_proof_2}
\end{align}
where in the first and fourth lines we use the second condition (\ref{eq:mono3_norm}) and the first condition (\ref{eq:mono1_norm}). This completes the proof of (\ref{eq:mono4_1}) and (\ref{eq:mono4_logN}).
The condition (\ref{eq:mono4_2}) is also satisfied, because trace norm is a convex function due to the triangle inequality $ \norm{A}+\norm{B}\geq \norm{A+B}$, and we have
\begin{align}
 \norm{\rho_{AB}^{\sT_{A}}} &\geq \sum_\mu r_\mu \norm{[\rho_{AB}(\mu)]^{\sT_{A}}}\\
 &\geq  \norm{\sum_\mu r_\mu [\rho_{AB}(\mu)]^{\sT_{A}}}.
\end{align}
\hfill $\blacksquare$


In Appendix~\ref{sec:bosonic_transpose}, we use the relation between the bosonic partial transpose~\cite{Eisler2015} and the fermionic partial transpose (\ref{eq:fermion_pt}) to show that the   A, B, and D properties also hold for the bosonic partial transpose of density matrices in fermionic systems and explain why C does not hold.

\emph{Note for bosonic partial transpose--} 
In general, LOCC operations can be used to prepare separable states with zero negativity. There also exists  a set of inseparable states which have zero negativity, but since they are inseparable they cannot be prepared  using solely LOCC operations. This phenomenon is known as bound entanglement~\cite{Horodecki1998} and suggests that there are other natural restricted classes of operations aside from LOCC.
Combining both classes of operators leads to a general class of positive partial transpose preserving operations (PPT operations) which have the property that they map the set of positive partial transpose states into itself.
In bosonic partial transpose, the monotonicity under PPT-operations has been shown in Refs.~\cite{Vidal2002,Plenio2005}.

 \subsection{Continuity}
 The entanglement measure should be continuous.
The Hilbert-Schmidt distance between two density matrices $\rho_1$ and $\rho_2$ is defined by
 \begin{align}
 D_{H S} (\rho_1, \rho_2) = \Tr[(\rho_1-\rho_2)^2 ] 
 \end{align}  
and is used as a measure of proximity of two states in the Hilbert space.
 It is required that if the Hilbert-Schmidt distance between two states vanishes, the difference between their entanglement should also go towards zero. There is not much to prove here, since the partial transpose is a linear operation and the negativity is defined in terms of a one-norm which is algebraically a continuous function.

 \subsection{Computability}
  The entanglement measure should be efficiently computable
for every state. For a generic density matrix represented in the occupation-number basis, it is straightforward to implement the transformation rule (\ref{eq:app_f_21}).  For Gaussian states including thermal states of quadratic Hamiltonians and reduced density matrices by partial tracing the ground state of such Hamiltonians, there is an efficient method (which scales linearly with the system size) to compute the negativity in terms of single-particle correlation function (covariance matrix), as we have shown in Ref.~\cite{Shap_pTR}.
Furthermore, the fermionic partial transpose could in principle be implemented in the matrix product state representation similar to what is done for the bosonic partial transpose~\cite{Ruggiero_1,Ruggiero_2}.


\section{Canonical examples of entangled states}
\label{sec:examples}

In this section, we give examples of bipartite and tripartite entangled states in fermionic systems.
We compare the results of bosonic and fermionic partial transpose in these examples and discuss the origin of possible differences. 

We construct entangled states of local fermionic modes in terms of ground states of some Hamiltonians, where we generate entanglement between two fermionic sites by adding a hopping (tunneling) term between them. We also compare fermionic states with the corresponding qubit states whenever such correspondence exists.  For clarity, the states of qubits are denoted by $\ket{\uparrow}$ and 
$\ket{\downarrow}$. We map the fermion operators $f_j$ and $f_j^\dag$, $ j=1,\cdots,N$ to those qubits by 
the Jordan-Wigner transformation
$S_j^-= \exp(i\pi \sum_{l<j} f^\dag_l f_l) f_l$
and
$S_j^+= \exp(-i\pi \sum_{l<j} f^\dag_l f_l) f_l^\dag$,
where $S^x_j=(S^+_j+S^-_j)/2$ and $S^z_j= S^+_j S^-_j -1/2$ are qubit (spin-$1/2$) operators.

\subsection{Bipartite systems}

Here, we consider the entangled states of two local fermionic modes. As we will see, the entangled states of two complex fermionic modes shares a lot of similarities with the states of two qubits. In contrast, the entangled states of two Majorana modes has no analog in two-qubit systems.

\subsubsection{Two complex fermions}

Let $f_1 \in {\cal G}(\Hi^A)$ and $f_2 \in {\cal G}(\Hi^B)$ be two fermionic sites.
In order to get an entangled state of two complex fermionic modes, we consider a simple hopping Hamiltonian,
\begin{align}
H=  \Delta (f_1^\dag f_2 + f_2^\dag f_1),
  \label{two mode Hamiltonian}
\end{align}
where $\Delta$ is the hopping amplitude.
The ground state of this Hamiltonian is given by
\begin{align}
\ket{\Psi_s}_f &=\frac{1}{\sqrt{2}}(f_1^\dag-f_2^\dag)\ket{0}.
\end{align}
This is a maximally entangled state and the logarithmic negativity is ${\cal E}=\log 2$ using either bosonic or 
fermionic partial transpose. The corresponding state in a system of two qubits is a spin-singlet,
\begin{align}
\ket{\Psi_s}_\text{qubit}= \frac{1}{\sqrt{2}} (\ket{\uparrow \downarrow} - \ket{\downarrow\uparrow}).
\end{align}

Now that we see the correspondence between the singlet state of qubits and two fermionic mode, it is instructive to consider Werner states which refer to a one-parameter set of mixed states,
\begin{align} \label{eq:Werner}
\rho= \frac{1}{4}(1-p) \mathbb{I}_A\otimes \mathbb{I}_B + p \ket{\Psi_s} \bra{\Psi_s}
\end{align}
where $0\leq p\leq 1$ is a real parameter which interpolates between a maximally mixed state (identity) and a maximally entangled state (singlet).

\begin{figure}
\includegraphics[scale=0.55]{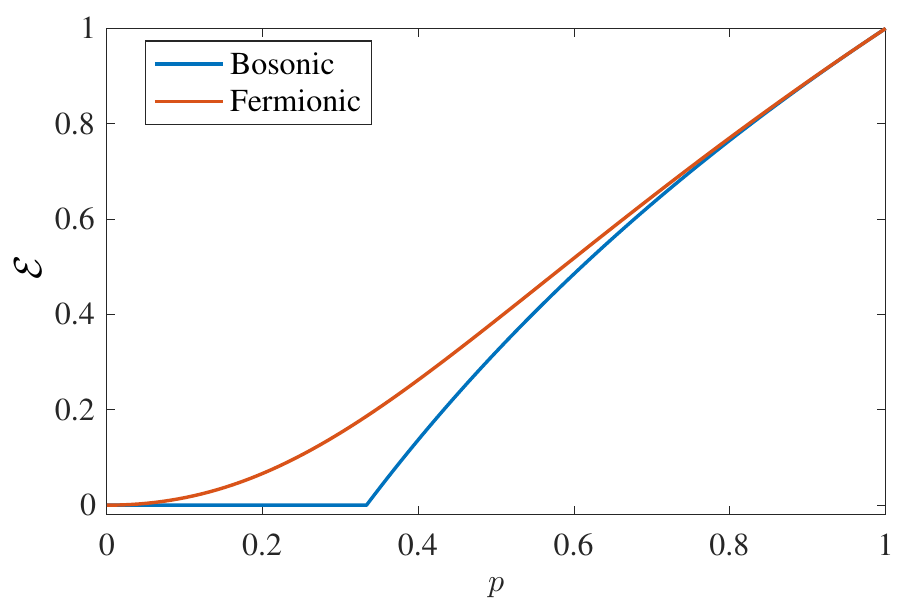}
\caption{\label{fig:two_cf} Comparison of entanglement negativity associated with fermionic and bosonic partial transpose for a Werner state (\ref{eq:Werner}).}
\end{figure}

It is easy to compute the bosonic negativity 
\begin{align} \label{eq:b_cf}
{\cal E}_b(\rho)=\log \left( \frac{3}{4} (1 + p) +\frac{1}{4} |1 - 3 p| \right).
\end{align}
On the other hand, the Werner state can be considered as a valid density matrix of fermions and so, the fermionic negativity is found to be
\begin{align} \label{eq:f_cf}
{\cal E}(\rho)=\log \left( \frac{1}{2}(1 + p) + \frac{1}{2}\sqrt{5 p^2 - 2 p + 1}  \right).
\end{align}
The bosonic (\ref{eq:b_cf}) and fermionic (\ref{eq:f_cf})  negativities are compared in Fig.~\ref{fig:two_cf}. We note that these two quantities approach each other as $p$ goes to zero or one and they are tangent at these two extreme points. Also, we observe that ${\cal E}\sim p^2$ in the regime $p\ll 1$, i.e.~the lowest order term in the fermionic negativity grows quadratically as we depart from a separable state. The quadratic behavior is rather generic (see also (\ref{eq:neg_generic_2cf})) and a basis of our proof of Theorem~\ref{thm:bisep}
 in the next section.

A notable difference between the bosonic and fermionic negativities is that ${\cal E}$ only vanishes at $p=0$, whereas ${\cal E}_b$ remains zero as long as $p\leq 1/3$. 
This difference can be understood as a consequence of the fermion-number parity
constraint on the density matrices in the fermionic Hilbert subspaces. The
Werner states  can be written in the occupation-number basis
$\{\ket{00},\ket{10},\ket{01},\ket{11}\}$ as 
\begin{align} 
\rho= \frac{1}{4}
\left( \begin{array}{cccc}
\frac{1-p}{4} & 0 & 0 & 0\\
0 & \frac{1+p}{4} & \frac{-p}{2} & 0 \\
0& \frac{-p}{2} & \frac{1+p}{4} & 0 \\
0 & 0 & 0 & \frac{1-p}{4} 
\end{array} \right).
\end{align}
The density matrix can be decomposed~\cite{Ben-Aryeh2015} as follows
\begin{align}
\rho=&\frac{p}{2} \sum_{i=x,y,z}  {\Big [} \frac{(\I-\sigma^i)_A}{2} \otimes \frac{(\I+ \sigma^i)_B}{2}
 \nonumber \\ 
&\ \ \ \ \ \ \ \ \ \ +
\frac{(\I+\sigma^i)_{A}}{2} \otimes \frac{(\I- \sigma^i)_{B}}{2} {\Big ]} \nonumber \\ 
& +  \left( \frac{1-3p}{4} \right)  \I_{A}\otimes \I_{B},
\end{align}
in terms of local operators represented by Pauli matrices in the occupation number basis. If we consider it as a bosonic system of qubits, each term is a product state since the operators are valid local density matrices. Therefore, the above decomposition shows that for $p\leq 1/3$ all the coefficients are positive and the density matrix is separable in the bosonic formalism.
However, the density matrices $\I\pm\sigma^x=1+f^\dag f \pm f\pm f^\dag$ (similarly, $\I\pm\sigma^y$) are not  legitimate density matrices of fermions, because they violate the fermion-number parity symmetry and the above decomposition implies that the density matrix is never separable in the fermionic formalism. Therefore, the separability criterion given by the bosonic partial transpose fails to address the separability of fermionic states correctly.

\subsubsection{Two Majorana fermions}
\label{sec:Thm1}
\begin{figure}
\includegraphics[scale=0.75]{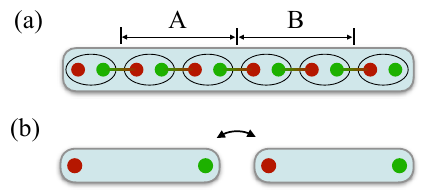}
\caption{\label{fig:2majorana} Hamiltonian of two coupled Majorana modes, (a) as in the reduced density matrix of two adjacent intervals on a single chain and (b) two coupled chains.}
\end{figure}

 Let $c_1 \in {\cal G}(\Hi^A)$ and $c_2 \in {\cal G}(\Hi^B)$ define two Majorana modes.
We consider a tunneling Hamiltonian between two Majorana fermion sites,
\begin{align} \label{eq:H_2majorana}
H=- i\Delta c_1 c_2.
\end{align}
The density matrix associated with the ground state manifold of this Hamiltonian effectively describes the reduced density matrix of  two adjacent intervals on a single Kitaev Majorana chain, where the reduced density matrix can be represented in terms of two edge Majorana fermions $c_1$ and $c_2$ at the interface between the two intervals~\cite{Shap_pTR} (Fig.~\ref{fig:2majorana}(a)). Furthermore,
the Hamiltonian (\ref{eq:H_2majorana}) can be physically related to the low energy modes of two coupled Majorana chains as shown in Fig.~\ref{fig:2majorana}(b).

In order to compute the state density matrix and the corresponding partial transpose, we use a description in which the two coupled Majorana sites are  part of two complex fermionic modes and represent the density matrix in a basis of two-fermion modes. We combine $c_1$ and $c_2$ with the two unentangled Majorana modes $c_3$ and $c_4$ (which can be physically thought of as the zero modes at the far end on each interval or chain shown in Fig.~\ref{fig:2majorana}(a) or (b)) to construct the complex fermions $f_1=(c_1+ic_3)/2$ and $f_2=(c_2+i c_4)/2$, where $f_1 \in {\cal G}(\Hi^A)$ and $f_2 \in {\cal G}(\Hi^B)$. The crucial point is that the entanglement does not depend on the representation.
 The density matrix can then be expressed in the occupation-number basis $\{\ket{00},\ket{10},\ket{01},\ket{11}\}$ as in
\begin{align} \label{eq:rho_Kitaev}
\rho= \frac{1}{4} ( 1- i c_1 c_2)=
 \frac{1}{4}
\left( \begin{array}{cccc}
1 & 0 & 0 & 1\\
0 & 1& 1& 0 \\
0& 1 & 1 & 0 \\
1& 0 & 0 & 1 
\end{array} \right).
\end{align}
With a little bit of algebra, one can see that ${\cal E}(\rho)=\log\sqrt{2}$ while the bosonic negativity vanishes. So, let us get some insights by checking the separability criterion. The density matrix can be expanded~\cite{Ben-Aryeh2015} in the following form
\begin{align} \label{eq:Kitaev_decomp}
\rho=& \frac{1}{8} (\I+\sigma^x)_{A}\otimes (\I +\sigma^x)_{B} \nonumber \\
&+  \frac{1}{8} (\I-\sigma^x)_{A}\otimes (\I -\sigma^x)_{B}.
\end{align}
We should note that $\rho$ is separable when it is viewed as density matrix of two qubits in the bosonic formalism. However, as also mentioned for the Werner states, $\I\pm\sigma^x=1+f^\dag f \pm f\pm f^\dag$ are not density matrices of fermions, and the state is not separable in the fermionic formalism. This observation means that the density matrix (\ref{eq:rho_Kitaev}) is separable from point of view of bosonic partial transpose while it is inseparable (entangled) from the point of view of fermionic partial transpose. 

Let us conclude this part with a general remark about a system of two fermionic modes. The most general form of single-mode fermionic density matrix allowed by fermion-number parity symmetry is
\begin{align} \label{eq:1f_separable}
\rho= \frac{\I+ \alpha \sigma^z}{2},
\end{align}
where $|\alpha|\leq 1$ so that $\rho$ is positive definite.
Taking two fermionic modes and constructing the full density matrix by tensor product,  the resulting matrix contains only diagonal elements and hence, it is evident that a given fermionic density matrix is not separable (from fermionic point of view) unless all off-diagonal elements are zero. 
Moreover, we find that vanishing logarithmic negativity is a necessary and sufficient condition for separability of a two-fermion density matrix.  Therefore, we can put forward the following theorem:

\begin{thm}
A two-fermion mixed state $\rho$ is separable if and only if ${\cal N}(\rho)=0$.
\label{thm:2f_sep}
\end{thm}

The necessary condition is already evident (see Sec.~\ref{sec:separability}).
The sufficient condition follows immediately from the remark that  any inseparable state has to have off-diagonal elements (see Appendix~\ref{sec:2f_proof} for details). 


\subsection{Tripartite systems}

In this part, we present analogs of tripartite entangled states for three local fermionic modes. 
As we see below, the W state is very similar in qubit and fermionic systems, while there are some differences between the Greenberger-Horne-Zeilinger  (GHZ) states of qubits and fermions.

\subsubsection{Three complex fermions}
\label{sec:Three complex fermions}

\begin{figure}
\includegraphics[scale=0.65]{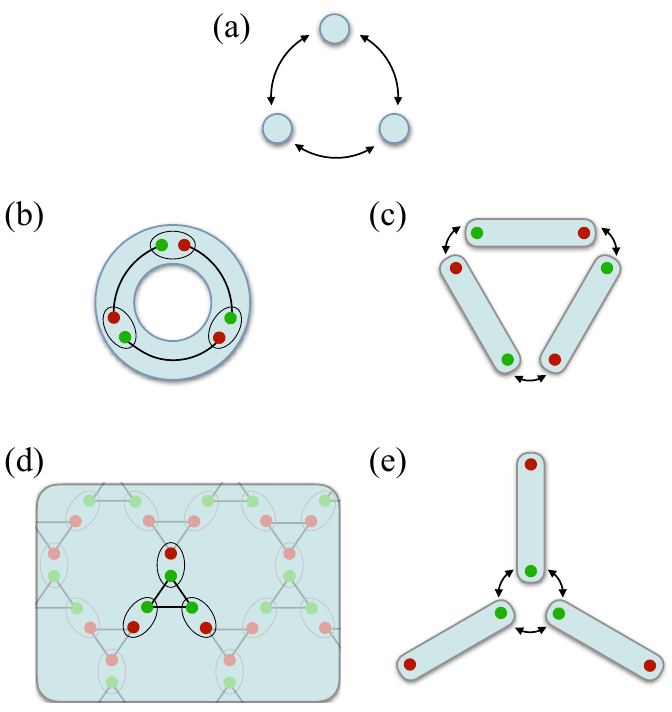}
\caption{\label{fig:fstates} Representation of Hamiltonians with entangled ground states. (a) The W state of complex fermions, where each blue circle indicates a complex fermionic mode. The GHZ state of complex fermions can be realized by (b) three-site Kitaev Majorana chain, or (c) three coupled Kitaev Majorana chains in the $\Delta$ geometry. Tri-partite entangled state of Majorana modes is obtained by (d) a subsystem of Kagome lattice with three sites, or (e)  three coupled Kitaev Majorana chains in the $Y$ geometry. In panels (c) and (e), each blue rod is a Kitaev Majorana chain and the smaller circles within each chain represent Majorana end-modes. Arrows in (a), (c) and (e) and solid lines in (b) and (d)  represent quadratic tunneling terms.}
\end{figure}

A W state of three qubits is given by
\begin{align} \label{eq:Wqubit}
\ket{W}_{\text{qubit}}= \frac{1}{\sqrt{3}} (\ket{\uparrow\downarrow\downarrow}+\ket{\downarrow\uparrow\downarrow}+\ket{\downarrow\downarrow\uparrow}),
\end{align}
This is the original symmetric W state, which is unique for distributing equal
entanglement between any two parties.
This state corresponds to the ground state of the Hamiltonian
\begin{align} \label{eq:3xy}
H= &-\frac{1}{2} \sum_{j=1}^3 [S^{x}_{j+1} S^{x}_{j} +S^{y}_{j+1} S^{y}_{j} ],
\end{align}
where $S^\alpha_4\equiv S^\alpha_1$.

Let $f_1 \in {\cal G}(\Hi^A)$, $f_2 \in {\cal G}(\Hi^B)$, and $f_3 \in {\cal G}(\Hi^C)$ be three local fermionic modes owned by three parties $A$, $B$, and $C$, respectively.
We obtain the W state of fermions from the ground state of
   the following Hamiltonian 
\begin{align}
H= - \sum_{j=1}^3 [f_{j+1}^\dag f_j + \text{H.c.}],
\end{align}
which is equivalent to (\ref{eq:3xy}) via a Jordan-Wigner transformation. 
Here $f_{4}\equiv f_1$ (see Fig.~\ref{fig:fstates}(a), where each term is shown as a bond between two sites). The ground state is given by
\begin{align} \label{eq:Wf}
\ket{W}_f=\frac{1}{\sqrt{3}} (f_1^\dag+f_2^\dag+f_3^\dag)\ket{0},
\end{align}
which has an identical form to the W state of qubits (\ref{eq:Wqubit}). Furthermore, the entanglement negativities are identical for both bosonic and fermionic partial transpose, and given by
\begin{align}
{\cal E}_{A(BC)} &= \log \tr |\rho^{\sT_A}| =\log \left(1+\frac{2}{3}\sqrt{2} \right),\\
{\cal E}_{AB} &= \log \tr |\rho_{AB}^{\sT_A}| =\log \left( \frac{2+\sqrt{5}}{3} \right),
\end{align}
where $\rho=\ket{W}_f \bra{W}$ and $\rho_{AB}$ is the reduced density matrix after tracing out $\Hi^C$, $\rho_{AB}= \tr_C (\rho)$.

The GHZ state of three qubits is defined by,
\begin{align} \label{eq:GHZqubit1}
\ket{GHZ}_\text{qubit}= \frac{1}{\sqrt{2}} (\ket{\downarrow\downarrow\downarrow}+\ket{\uparrow\uparrow\uparrow})
\end{align}
which resembles a superposition of the ordered phase of the Ising chain of three spins,
\begin{align}
H= -\sum_{j=1}^3 S^x_{j+1} S^x_{j}
\end{align}
where $S^\alpha_4\equiv S^\alpha_1$.

This suggests that we should consider the fermionized Ising chain Hamiltonian (via the Jordan-Wigner transformation),
\begin{align} \label{eq:Kitaev_H}
H= - \sum_{i=1}^3 [f_{i+1}^\dag f_i+f_{i+1}^\dag f_i^\dag + \text{H.c.}]
\end{align}
that is basically the Kitaev Majorana chain of three sites (Fig.~\ref{fig:fstates}(b)) or three coupled Majorana chains in $\Delta$ geometry as shown in Fig.~\ref{fig:fstates}(c).
Then, the ground state of this Hamiltonian defines a GHZ state of fermions,
\begin{align} \label{eq:GHZf}
\ket{GHZ}_f=\frac{1}{2} (f_1^\dag+f_2^\dag+f_3^\dag + f_1^\dag f_2^\dag f_3^\dag)\ket{0}.
\end{align}
At first look, one may not see any similarity between this state and the GHZ state of qubits (\ref{eq:GHZqubit1}) upon identifying the Hilbert spaces of qubits and fermionic sites. This is because the fermionized GHZ state of qubits $(1+f_1^\dag f_2^\dag f_3^\dag)\ket{0}$ is not a legitimate fermionic state as it is a superposition of two basis vectors with different fermion-number parities. However, if we rewrite the qubit state in terms of eigenstates of $S^x$ or $S^y$, we will get
\begin{align} 
\ket{GHZ_x}_\text{qubit}&= \frac{1}{\sqrt{2}} (\ket{\downarrow_x\downarrow_x\downarrow_x}+\ket{\uparrow_x\uparrow_x\uparrow_x}) \\
&= \frac{1}{2} (\ket{\uparrow\downarrow\downarrow}+\ket{\downarrow\uparrow\downarrow}+\ket{\downarrow\downarrow\uparrow}+\ket{\uparrow\uparrow\uparrow})
\label{eq:GHZqubit2}
\end{align}
where $\ket{\uparrow_x}=(\ket{\uparrow}+\ket{\downarrow})/\sqrt{2}$  and $\ket{\downarrow_x}=(\ket{\uparrow}-\ket{\downarrow})/\sqrt{2}$ are eigenstates of $S^x$. Each term in this expression is now in one-to-one correspondence with the fermionic GHZ (\ref{eq:GHZf}). This type of GHZ state can be generated by applying the braiding operators of the Temperley-Lieb algebra to separable basis states~\cite{Solomon2010}. We should note that although the  bosonic GHZ states (\ref{eq:GHZqubit1}) and (\ref{eq:GHZqubit2}) are unitarily equivalent, the entanglement within each pair of parties is different after a projective measurement of the third qubit in the standard up/down basis. In other words, measuring the $A$ qubit in $\{\ket{\uparrow}, \ket{\downarrow} \}$-basis gives a product state of the qubits $B$ and $C$ in  (\ref{eq:GHZqubit1}), while the same process on the state (\ref{eq:GHZqubit2}) gives a maximally entangled Bell state of the qubits $B$ and $C$.

The entanglement negativity of
the $\ket{GHZ}_f$ state (\ref{eq:GHZf}) with respect to each party is
\begin{align}
{\cal E}_{A(BC)}=\log 2
\end{align}
which is identical for both bosonic and fermionic partial transpose. In the fermionic system, this amount of entanglement corresponds to two Majorana bonds between each party and the other two (Fig.~\ref{fig:fstates}(b)).
The reduced density matrix obtained after tracing out one fermionic mode is identical to (\ref{eq:rho_Kitaev}). This should have been clear since the Hamiltonian (\ref{eq:Kitaev_H}) describes a Kitaev Majorana chain.
Hence, all three reduced entanglement negativities are zero using the bosonic partial transpose, while ${\cal E}_{AB}=\log \sqrt{2}$ using the fermionic partial transpose.

\subsubsection{Three Majorana fermions}

Now that we learn how to construct entangled states of three complex fermionic modes by writing a proper Hamiltonian, let us try it for the real fermions.
We should note that there is no difference between the GHZ and W states for real fermions using this approach, because there is only one type of coupling term between Majorana operators as they are self-adjoint.

Let $c_1 \in {\cal G}(\Hi^A)$, $c_2 \in {\cal G}(\Hi^B)$, and $c_3 \in {\cal G}(\Hi^C)$ be three Majorana modes owned by three parties $A$, $B$, and $C$, respectively.
A tripartite entangled state is obtained from the ground state of the Hamiltonian
\begin{align}
H=i \sum_{j=1}^3  c_{j+1} c_{j},
\end{align}
where $c_{4}\equiv c_1$. For example, the density matrix associated with the ground state space of this Hamiltonian can be identified as a reduced density matrix of a subsystem of three sites embedded in a lattice model as shown in Fig.~\ref{fig:fstates}(d) (for a Kagome lattice), or it may describe the physics of three coupled Majorana chains in a $Y$ geometry depicted in Fig.~\ref{fig:fstates}(e).
We use the same idea as in the case of two Majorana fermions by considering extra Majorana modes $b_j$ to pair up with $c_j$'s and construct complex fermions (shown as red circles in Fig.~\ref{fig:fstates}(d) and (e)). 
Equivalently, we observe that $b_j$ Majorana operators do not appear in the Hamiltonian and we get four-fold ground state degeneracy. Then, the state density matrix of $c_j$ Majorana modes can be obtained by summing over the density matrices of the degenerate ground states with equal weights. The resulting state is given by
\begin{align} 
\rho_{ABC}=\frac{1}{8} [1- \frac{i}{\sqrt{3}}  \sum_{j=1}^3  c_{j+1} c_{j}].
\end{align}
We should note that similar to the case of two entangled Majorana modes, here the state is described by a density matrix. The fermionic entanglement negativity with respect to each party becomes
\begin{align}
{\cal E}_{A(BC)}=\log  \sqrt{\frac{5}{3}}.
\end{align}
Moreover, the reduced density matrix is found by
\begin{align}
\rho_{AB}=\Tr_{C}(\rho_{ABC})=\frac{1}{4} (1+\frac{i}{\sqrt{3}}c_1 c_2),
\end{align}
which yields the fermionic negativity of
\begin{align}
  {\cal E}_{AB}=\log \frac{2}{\sqrt{3}}.
\end{align}
It is worth noting that the bosonic entanglement negativity of either density
matrices $\rho_{ABC}$ or $\rho_{AB}$ vanishes~\cite{Shap_pTR}. 
This is due to the fact that there is no analog of Majorana modes in qubit systems and these states look like a separable state in the occupation number basis according to bosonic formalism.

In the two subsequent sections, we use the machinery of fermionic partial transpose and fermion-number parity to present some ideas towards classifying pure and mixed states of a system with two and three fermionic modes.

\section{Classification of two-fermion states}
\label{sec:class two}

Here, the system of interest consists of two local fermionic modes $f_1 \in {\cal G}(\Hi^A)$ and $f_2 \in {\cal G}(\Hi^B)$, each belonging to the $A$ and $B$ parties, respectively. 

\subsection{Pure states}

Any state of a two-fermion system can be written in either of the following forms
\begin{align} \label{eq:2f_pure_class}
\ket{\Psi_+} &= (\lambda_0 +\lambda_1 f_1^\dag f_2^\dag)\ket{0}_f,  \nonumber \\
\ket{\Psi_-} &= (\lambda_0 f_1^\dag  +\lambda_1 f_2^\dag)\ket{0}_f,
\end{align}
associated with $(-1)^F\ket{\Psi_\pm}=\pm\ket{\Psi_\pm}$, where $\lambda_i \in \mathbb{C}$ and $|\lambda_0|^2+|\lambda_1|^2=1$.
The situation here is quite clear and any bipartite entanglement measure can distinguish two classes of states: entangled vs. unentangled. For instance, the entanglement negativity yields
\begin{align}
{\cal N}(\rho) = |\lambda_0 \lambda_1|,
\end{align}
where $\rho=\ket{\Psi}\bra{\Psi}$ for either sectors.

\subsection{Mixed states}

There are two classes of states: separable and inseparable. As we discussed earlier, the separability criterion of two fermions is equivalent to 
vanishing logarithmic negativity (Theorem~\ref{thm:2f_sep} in Sec.~\ref{sec:examples}).
Hence, these two classes can be unambiguously distinguished by the logarithmic negativity. In other words, ${\cal E}(\rho_\text{sep})=0$ while ${\cal E}(\rho_\text{insep})>0$.

\section{Classification of three-fermion states}
\label{sec:class three}

Here, we consider a system of three local fermionic modes $f_1 \in {\cal G}(\Hi^A)$, $f_2 \in {\cal G}(\Hi^B)$, and $f_3 \in {\cal G}(\Hi^C)$ owned by the $A$, $B$, and $C$ parties, respectively. 
The game is to find good entanglement measures which diagnose all possible classes of states.

There have been several attempts at classifying the pure states of three-qubit systems~\cite{Dur2000,Acin2000,Ou2007,Solomon2012}. As we will see, the methods of Refs.~\cite{Dur2000,Ou2007,Solomon2012} can be applied to fermionic systems.
However, Ref.~\cite{Acin2000} introduces a method in which a given state is brought to
a certain canonical form by unitary transformations and a generalized Schmidt decomposition  are derived for those canonical states. This approach is not applicable to fermionic systems as the mentioned unitary transformation is not a valid operation in the fermionic formalism because of violating the fermion-number parity symmetry.

The classification of mixed states of three qubits is much more complicated and there is no universal framework which addresses all possible states~\cite{Dur1999,*Dur2000_mixed,Acin2001,Sabin2008,Eltschka2012}.
Due to numerous possibilities of three-qubit mixed states, previous studies have tried to devise technologies to distinguish certain families of density matrices which obey a certain form. For instance, Dur~\emph{et.~al.}~\cite{Dur1999,*Dur2000_mixed} used negativities to classify superposition of GHZ density matrices associated with a set of orthonormal GHZ states and Acin~\emph{et.~al.}~\cite{Acin2001} introduced entanglement witnesses to distinguish mixed states of GHZ and W states with positive coefficients. Interestingly, Ref.~\cite{Acin2001} finds that the W-type mixed states form a subspace of finite volume as opposed to the W-type pure states which occupies a measure-zero subspace in a parameterized Hilbert space.  Reference~\cite{Eltschka2012} also provided a classification for a family of GHZ symmetric mixed states (parametrized in a Euclidean space with the Hilbert-Schmidt metric) based on the invariance of entanglement properties under general local operations.




\subsection{Pure states}

Due to fermion-number parity constraint, generic pure states of three fermionic modes are
\begin{align} \label{eq:pure_3f_1}
\ket{\Psi_+} &= [\lambda_0 + \lambda_1 f_{1}^\dag f_{2}^\dag+ (\lambda_2 f_{2}^\dag  + \lambda_3 f_{1}^\dag) f_{3}^\dag] \ket{0}_f,  \\
\ket{\Psi_-} &= [ (\lambda_0 + \lambda_1 f_{1}^\dag f_2^\dag) f_3^\dag +\lambda_2 f_{2}^\dag+ \lambda_3 f_{1}^\dag ] \ket{0}_f,
\label{eq:pure_3f_2}
\end{align}
corresponding to even and odd sectors of the Hilbert space $\Hi^A\otimes \Hi^B\otimes \Hi^C$. We assume that the states are normalized $\sum_i |\lambda_i|^2=1$.
From this representation of states, we can categorize pure states into four types:
\begin{enumerate}
\item
\emph{Separable state:}
Any single term out of the four terms is obviously a product state and hence separable.

\item
\emph{Bi-separable state:}
Any two terms out of four is a bi-separable state, i.e., one of the parties is not entangled to the other two. 

\item
\emph{W state:}
Any combination of three terms in the above pure states (\ref{eq:pure_3f_1}) or (\ref{eq:pure_3f_2}). 

\item
\emph{GHZ state:}
A state with all four terms present.
\end{enumerate}

The entanglement of each party to the rest of system can be measured by 
\begin{align}
{\cal N}_{A(BC)}&=\frac{\Tr |\rho^{\sT_A}| -1}{2} \nonumber \\
&=  \left[(|\lambda_0|^2+|\lambda_2|^2)(|\lambda_1|^2+|\lambda_3|^2)\right]^{1/2} 
\end{align}
or
\begin{align} 
{\cal E}_{A(BC)}&= \log \Tr |\rho^{\sT_A}| \nonumber \\
&= \log \left[1+2 ((|\lambda_0|^2+|\lambda_2|^2)(|\lambda_1|^2+|\lambda_3|^2))^{1/2}  \right]
\end{align}
where $\rho=\ket{\Psi}\bra{\Psi}$. Note that ${\cal E}_{A(BC)}$ is also identical to $1/2$-R\'enyi entropy of party $A$'s reduced density matrix $\rho_{A}=\Tr_{BC}(\rho)
$, i.e., $S_{1/2}(\rho_A)=2\log \Tr \rho_{A}^{1/2}$,  since $\rho$ is a pure state. We can define ${\cal E}_{B(AC)}$ and ${\cal E}_{C(AB)}$ (or ${\cal N}_{B(AC)}$ and ${\cal N}_{C(AB)}$) similarly. So far, we realize that the tuple  $({\cal N}_{A(BC)},{\cal N}_{B(AC)},{\cal N}_{C(AB)})$ or $({\cal E}_{A(BC)},{\cal E}_{B(AC)},{\cal E}_{C(AB)})$ can distinguish separable and bi-separable states from each other and from W and GHZ states. In fact, any type of bipartite entanglement entropy works equally well for this purpose~\cite{Dur2000}. However, they cannot differentiate between W and GHZ states.
To this end, we introduce a product
\begin{align}
J_{ABC}={\cal N}_{AB,e} {\cal N}_{AB,o},  
\end{align}
of the reduced entanglement negativities in the even and odd fermion-number parity sectors by
\begin{align}
{\cal N}_{AB,e} &= \frac{ \Tr |\rho_{AB,e}^{\sT_A}|-1}{2}, \\
{\cal N}_{AB,o} &=  \frac{ \Tr |\rho_{AB,o}^{\sT_A}|-1}{2},  
\end{align}
where 
\begin{align}
\rho_{AB,s}= \frac{1}{p_s} \Tr_C (P_s \rho P_s),
\end{align}
are projected density matrices onto even and odd sectors $s=e,o$, where $p_s=\Tr(\rho_{AB,s})$ is the normalization factor. The projection operators are
\begin{align}
P_e &= \frac{1}{2}[1+ (-1)^{f_3^\dag f_3}] = \vac, \\
P_o &= \frac{1}{2}[1- (-1)^{f_3^\dag f_3}] = f_3^\dag \vac f_3.
\end{align} 
This quantity is a special indicator of GHZ-type states. Recall that GHZ-states correspond to ground states of a Majorana chain (\ref{eq:Kitaev_H}) which is a symmetry protected topological (SPT) phase protected by the $\Z_2$ fermion-number parity symmetry. The two projected negativities ${\cal N}_{AB,e}$ and ${\cal N}_{AB,o}$ can be viewed as SPT entanglements which detect the topological phase of Majorana chains~\cite{Marvian}.
 We should note that the choice of partitioning into $AB$ and $C$ does not matter for the product ${\cal N}_{AB,e} {\cal N}_{AB,o}$ and any partition which separates two parties from the other works equally well.
For $AB$ and $C$ partitions, we have
\begin{align}
{\cal N}_{AB,e} &=\frac{|\lambda_0 \lambda_1|}{|\lambda_0|^2+|\lambda_1|^2}, \\
{\cal N}_{AB,o} &=\frac{|\lambda_2 \lambda_3|}{|\lambda_2|^2+|\lambda_3|^2}.
\end{align}
It is easy to see that for a GHZ state we have $J_{ABC} >0$, while for a W state we get $J_{ABC}=0$. Table~\ref{tab:tripartite_pure} summarizes how the three negativities with respect to each party along with the negativity of the reduced density matrix can fully determine which of the six possible classes a given pure state fits in.

\begin{table} [!]
\begin{center}
\begin{tabular}{l c c c c }
 Class & $\ {\cal N}_{A(BC)}\ $ & $\ {\cal N}_{B(AC)}\ $ & $\ {\cal N}_{C(AB)}\ $ & $\ J_{ABC}$  \\
\hline
\hline
$A- B - C$ & $0$ & $0$ & $0$ & $0$ \\
\hline
$A- BC$ & $0$ & $>0$ & $>0$ & $0$ \\
\hline
$B- AC$ & $>0$ & $0$ & $>0$ & $0$ \\
\hline
$C- AB$ & $>0$ & $>0$ & $0$ & $0$ \\
\hline
W & $>0$ & $>0$ & $>0$ & $0$ \\
\hline
GHZ & $>0$ & $>0$ & $>0$ & $>0$ \\
\hline
\hline
\end{tabular}
\caption{
\label{tab:tripartite_pure} Classification of  three-fermion pure states. Dashes represent separable parties.}
\end{center}
\end{table}

An alternative to the quantity $J_{ABC}$ is the so-called three-tangle~\cite{Coffman2000}, which can also be computed in terms of the Cayley hyper-determinant \cite{Solomon2012}.  
For a tripartite state of qubits
\begin{equation}\label{psi}
\ket{\Psi}=\sum_{i,j,k=0}^1 a_{ijk}
\ket{ijk}\; \; \; \;  (i,j,k=0,1),
\end{equation}
 the Cayley hyperdeterminant of the
coefficient hypermatrix $A=(a_{ijk})$ is defined by
\begin{eqnarray}
{\rm HDet}\, A &=& a_{000}^2 a_{111}^2 + a_{001}^2 a_{110}^2 +
a_{100}^2
a_{011}^2\nonumber\\
 &-& 2 [a_{000} a_{001} a_{110} a_{111} + a_{000} a_{010} a_{101} a_{111}
 \nonumber\\ &+& a_{000} a_{011} a_{100} a_{111}+a_{001} a_{010} a_{101} a_{110} \nonumber\\
 &+& a_{001} a_{011} a_{101} a_{100} +  a_{010} a_{011} a_{101} a_{100}
]\label{Det-A}\\
 &+& 4\left[a_{000} a_{011} a_{101} a_{110} +  a_{001} a_{010} a_{100}
 a_{111}\right].\nonumber
\end{eqnarray}
In fermionic systems, the three-tangle is simplified into
\begin{align}
\tau_{ABC}= {\rm HDet} A= 4 \prod_{j=0}^3 |\lambda_j|,
\end{align}
because of the fermion-number parity constraint on the fermionic states. 
 Notice that product of all coefficients also appears in the numerator of $J_{ABC}$
and hence, the three-tangle $\tau_{ABC}$ (hyper-determinant) is non-zero only for a GHZ state, similar to  qubit systems.
 
In addition, two other measures of tripartite entanglement were introduced in terms of negativities~\cite{Ou2007,Sabin2008}. The first entanglement measure is defined as a geometric mean of entanglement negativities~\cite{Sabin2008},
\begin{align}
N_{ABC}= ({\cal N}_{A(BC)} {\cal N}_{B(AC)} {\cal N}_{C(AB)})^{1/3}
\end{align}
where ${\cal N}_{A(BC)}, {\cal N}_{B(AC)}$, and ${\cal N}_{C(AB)}$ are the negativities obtained from $\rho^{\sT_A}$, $\rho^{\sT_B}$, and $\rho^{\sT_C}$ (\ref{eq:neg}), respectively. The corresponding value for fermionic states (\ref{eq:pure_3f_1}) and (\ref{eq:pure_3f_2}) is given by
\begin{align}
N_{ABC}= \prod_{i \neq j} (|\lambda_i|^2 + |\lambda_j^2|)^{1/6}.
\end{align}
Clearly, $N_{ABC}$ vanishes for a separable or bi-separable state and is non-zero only for tripartite W and GHZ states.
The other measure of tripartite entanglement is called three-$\pi$ entanglement~\cite{Ou2007,Sabin2008}. This entanglement measure makes use of the inequality
\begin{align}
{\cal N}_{AB}^2+{\cal N}_{AC}^2 \leq {\cal N}_{A(BC)}^2,
\end{align}
in which ${\cal N}_{AB}= (\Tr|\rho_{AB}^{\sT_A}|-1)/2$ is computed for the reduced density matrix after tracing out $\Hi^C$ and similarly for ${\cal N}_{AC}$. The inequality becomes an equality for separable and bi-separable states.
Hence, an indicator of W and GHZ states can be constructed in terms of a `residual' entanglement with respect to each party
\begin{align}
\pi_A &= {\cal N}_{A(BC)}^2 - {\cal N}_{AB}^2-{\cal N}_{AC}^2, \\
\pi_B &= {\cal N}_{B(AC)}^2 - {\cal N}_{AB}^2-{\cal N}_{BC}^2, \\
\pi_C &= {\cal N}_{C(AB)}^2 - {\cal N}_{AC}^2-{\cal N}_{BC}^2, 
\end{align}
and the three-$\pi$ entanglement is given by
\begin{align}
\pi_{ABC}=\frac{\pi_A+\pi_B+\pi_C}{3}.
\end{align}
An important property of three-$\pi$ entanglement is that it can be applied to mixed-states, since it is fully defined in terms of negativities. A close form of $\pi_{ABC}$ for a generic state is bit lengthy. So, let us just mention the corresponding values 
\begin{align}
\pi_{ABC}^W= \frac{1}{9} (\sqrt{5}-1),
\end{align}
and
\begin{align}
\pi_{ABC}^{GHZ}=\frac{1}{4} (4\sqrt{2}-5) ,
\end{align}
for symmetric W (\ref{eq:Wf}) and GHZ state (\ref{eq:GHZf}), respectively. It is worth noting that $\pi_{ABC}$ of the W state is identical for bosonic and fermionic partial transpose, whereas it is different for the GHZ state ($\pi_{ABC}^{GHZ}=1/4$ for bosonic partial transpose) because the reduced negativities are different (c.f., \ref{sec:Three complex fermions}). Nevertheless, the fermionic three-$\pi$ entanglement satisfies $\pi_{ABC}^W< \pi_{ABC}^{GHZ}$, similar to the bosonic counterpart.

\begin{figure}
\includegraphics[scale=.57]{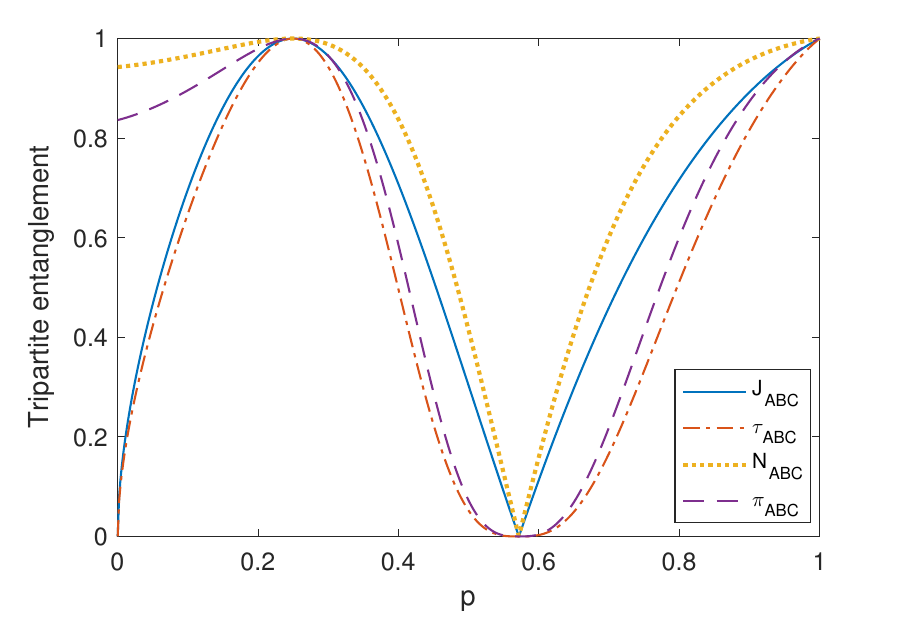}
\caption{\label{fig:tripartite} Comparison of various measures of tripartite entanglement in a one parameter fermionic state $\ket{\Psi_p}=\sqrt{p}\ket{GHZ}_f-\sqrt{1-p}\ket{W}_f$. All entanglements are normalized to give $1$ for a symmetric $\ket{GHZ}_f$ state.}
\end{figure}

Figure~\ref{fig:tripartite} compares the four measures of tripartite entanglement for a one-parameter family of states
\begin{align}
\ket{\Psi_p}=\sqrt{p}\ket{GHZ}_f-\sqrt{1-p}\ket{W}_f,
\end{align}
which interpolates between the symmetric GHZ and W states given by (\ref{eq:GHZf}) and (\ref{eq:Wf}), respectively.
We observe that at $p=4/7$ we get a separable state $\ket{\Psi_{p=4/7}}=f_1^\dag f_2^\dag f_3^\dag\ket{0}$, and all measures vanish. Furthermore, all measures are maximized at the GHZ type states, either at $p=1$ where $\ket{\Psi_{p=1}}=\ket{GHZ}_f$ or at $p=1/4$ where  $\ket{\Psi_{p=1/4}}=\frac{1}{2}(f_1^\dag f_2^\dag f_3^\dag-f_1^\dag-f_2^\dag-f_3^\dag)\ket{0}$. One major difference is that for W states $J_{ABC}$ and $\tau_{ABC}$ are zero, whereas $N_{ABC}$ and $\pi_{ABC}$ are not zero (but smaller than that of GHZ states).

\subsection{Mixed states}
\label{sec:thm2}

\begin{figure}
\includegraphics[scale=0.74]{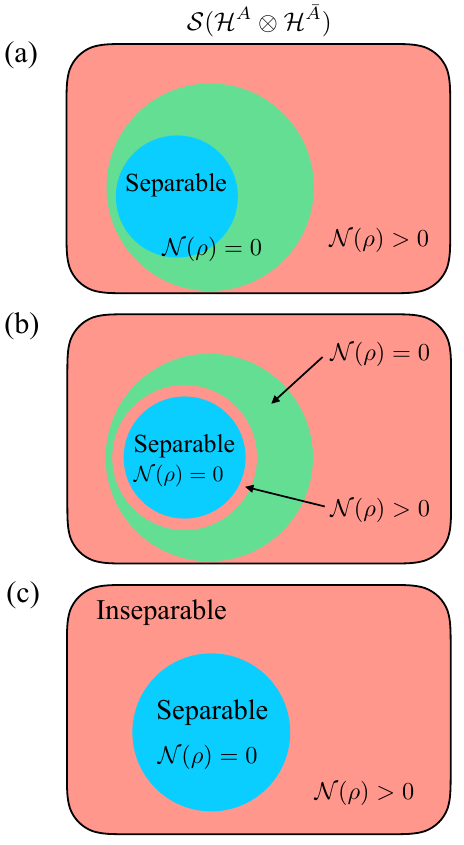}
\caption{\label{fig:sep} Schematic structure of separable and inseparable sets of states and values of the negativity in a bipartite system of one fermion $\Hi^A$ and a Fock space $\Hi^{\bar A}$ (discussed in Theorem~\ref{thm:bisep}).  (a) Separable states occupy a part of the zero-negativity subspace. 
(b) A situation in which there exists a subset of inseparable states with zero negativity away from separable states, that results in a contradiction (see the text). Therefore, the correct structure is given by (c), which means zero negativity is a necessary and sufficient condition for separability in this setup.}
\end{figure}

Here, we are dealing with a density matrix $\rho_{ABC}\in {\cal S}(\Hi^A\otimes\Hi^B\otimes\Hi^C)$.
We do not aim to classify all possible density matrices of three fermionic
modes, that is a daunting task beyond the scope of our paper.
Instead, we limit our discussion to
how the entanglement negativity with respect to each party
can constrain the form of the density matrix. For qubit systems, these constraints are found to be sufficient to fix the form of density matrix for certain families of states such as generalized Werner states and mixed orthonormal GHZ states~\cite{Dur1999,Dur2000_mixed}.
In what follows, we put forward a powerful theorem based on which we can build two important theorems. The latter theorems are particularly helpful in diagnosing separable and bi-separable mixed states of three fermions. 

\begin{thm}
Consider a density matrix $\rho \in {\cal S}(\Hi^A\otimes \Hi^{\bar A})$ where $\Hi^A$ contains one fermionic mode and $\Hi^{\bar A}$ is an arbitrary Fock space.
{$\rho$ can be decomposed as}
\begin{align} \label{eq:septhm}
\rho= \sum_i w_i \rho_{A,i} \otimes \rho_{\bar A,i}
\end{align}
where $w_i\geq 0$, if and only if ${\cal N}(\rho)=0$.
\label{thm:bisep}
\end{thm}

\noindent Proof:

The necessary condition is clearly true, given that  the property (\ref{eq:separable}) holds for fermionic partial transpose.
For the sufficient condition, in Appendix~\ref{sec:perturb_proof} we use perturbation theory to show that for any inseparable state in an immediate vicinity of separable states
$\rho=\rho_{\text{sep}} + \rho_{\text{off}}$, the negativity is non-zero to the leading order.
Here,
\begin{align}
\rho_{\text{sep}}= w_0 \vac \otimes \rho_0 + w_1 f_1^\dag \vac f_1 \otimes \rho_1 
\end{align}
is a separable part in which
$f_1\in {\cal G}(\Hi^A)$ and $\rho_1, \rho_2 \in {\cal S}(\Hi^{\bar A})$ are two fermionic density matrices, and
\begin{align}
\rho_{\text{off}} =  f_1^\dag \vac  \otimes \delta \rho +  \vac f_1   \otimes \delta \rho^\dag,
\end{align}
is an inseparable part where $\delta\rho \in {\cal G}_1(\Hi^{\bar A}) $ is a fermion-number parity odd operator. 

The rest of the proof is a consequence of the convexity of the zero-negativity states (Theorem~\ref{thm:PPT_convex}). Note that separable states form a convex set as shown in Fig.~\ref{fig:sep}(a). Our finding from perturbation theory implies that there exists at least a finite strip of states $S_1$  right outside the boundary of separable states (shown as a red strip in Fig.~\ref{fig:sep}(b)) where ${\cal N}(\rho)>0$. Let us assume that there exists a subset $S_2$ of states  outside this strip (green region) where the negativity vanishes ${\cal N}(\rho)=0$. This immediately contradicts with the convexity of the set of states with zero negativity. 
 Therefore, ${\cal N}(\rho)>0$ for all inseparable states (as depicted in Fig.~\ref{fig:sep}(c)).
\hfill $\blacksquare$


It is worth noting that a special case of Theorem~\ref{thm:bisep} is Theorem~\ref{thm:2f_sep} where $\Hi^{\bar A}$ consists of one fermionic mode. Another special case is applicable to the mixed state of three fermions, where $\Hi^{\bar A}=\Hi^B\otimes \Hi^C$ containing both $f_2$ and $f_3$ modes.

Let us recall that the negativity with respect to each party in a three-fermion system is defined by
\begin{align}
{\cal N}_{A(BC)}=\frac{ \Tr |\rho^{\sT_A}|-1}{2}
\end{align}
and similarly for ${\cal N}_{B(AC)}$ and ${\cal N}_{C(AB)}$.

\begin{cor}
{$\rho$ can be decomposed as}
\begin{align}
\rho= \sum_i w_i \rho_{A,i} \otimes \rho_{BC,i}
\end{align}
{where $w_i\geq 0$, if and only if ${\cal N}_{A(BC)}=0$. }
\end{cor}

As a result of the above corollary, we arrive at the following two theorems.

\begin{thm}
$\rho$ is fully separable,
\begin{align}
\rho_{}= \sum_i w_i\ \rho_{A,i} \otimes \rho_{B,i} \otimes \rho_{C,i},
\end{align}
{if and only if ${\cal N}_{A(BC)}={\cal N}_{B(AC)}={\cal N}_{C(AB)}=0$.}
\label{thm:sep_3f}
\end{thm}

\noindent Proof:

The necessary condition is a direct consequence of our result in Sec.~\ref{sec:separability}.
Moreover, because of Theorem~\ref{thm:bisep}, having ${\cal N}_{A(BC)}=0$ implies the following form for the density matrix
\begin{align}
\rho= \sum_j r_j \rho_{A,j} \otimes \rho_{BC,j}
\end{align}
where $r_j\geq 0$. A general form of the above bi-separable state can be written as
\begin{align}
\rho= w_0 \vac \otimes \rho_0 + w_1 f_1^\dag \vac f_1 \otimes \rho_1 
\end{align}
where $\rho_1, \rho_2 \in {\cal S}(\Hi^B\otimes \Hi^C)$ are two fermionic density matrices and $w_0+w_1=1$. Negativity with respect to $B$ is found by
\begin{align}
\Tr |\rho^{\sT_B}| &= w_0 \Tr |\rho^{\sT_B}_0| + w_1 \Tr |\rho^{\sT_B}_1|.
\end{align}
Having ${\cal N}_{B(AC)}=0$ implies that 
\begin{align}
w_0\Tr |\rho^{\sT_B}_0|+w_1\Tr |\rho^{\sT_B}_1|=1.
\end{align}
Since $\Tr |\rho^{\sT_B}_i|\geq 1$ and $w_i$ are positive coefficients which sum to one, the only way that the above equality holds is when $\Tr |\rho^{\sT_B}_i|=1$.
Using Theorem~\ref{thm:2f_sep} for two fermionic modes, we deduce that both $\rho_0$ and $\rho_1$ are separable and therefore, the sufficient condition of Theorem~\ref{thm:sep_3f} holds. A side remark here is that once ${\cal N}_{A(BC)}={\cal N}_{B(AC)}=0$, it is guaranteed that ${\cal N}_{C(AB)}=0$ as well.
\hfill $\blacksquare$

A straightforward consequence of this theorem is a useful criterion for bi-separability
\begin{thm}
{$\rho$ is bi-separable,}
\begin{align} \label{eq:rho_bisep}
\rho= \sum_i w_i \rho_{A,i} \otimes \rho_{BC,i},
\end{align}
{where at least one of $\rho_{BC,i}$ are inseparable, if and only if ${\cal N}_{A(BC)}=0$, but ${\cal N}_{B(AC)}\neq 0$ and ${\cal N}_{C(AB)}\neq 0$.}
\end{thm}

To appreciate the rich structure of three-fermion mixed states, we close this section by giving an example of a density matrix which is bi-separable (\ref{eq:rho_bisep}), but the reduced density matrix after tracing out $A$ is separable. This situation only arises in mixed states. Consider the following density matrix
\begin{align}
\rho=\frac{1}{2}( f_1^\dag \ket{\Psi_+}\bra{\Psi_+} f_1  + \ket{\Psi_-}\bra{\Psi_-}),
\end{align}
where 
\begin{align}
\ket{\Psi_\pm}=\frac{1}{\sqrt{1+|\alpha|^2}}(1\pm \alpha f_2^\dag f_3^\dag)\ket{0}.
\end{align}
This is a bi-separable state (\ref{eq:rho_bisep}) by construction and we have ${\cal N}_{A(BC)}=0$, ${\cal N}_{B(AC)}\neq 0$ and ${\cal N}_{C(AB)}\neq 0$.
However, the reduced density matrix of $BC$ is separable,
\begin{align}
\rho_{BC} &= \Tr_A(\rho) \nonumber \\
&= \frac{1}{2} (\ket{\Psi_+}\bra{\Psi_+}   + \ket{\Psi_-}\bra{\Psi_-}) \\
&=\frac{1}{{1+|\alpha|^2}} ( |\alpha^2| f_2^\dag f_3^\dag \vac f_3 f_2 + \vac),
\end{align}
which gives ${\cal N}_{BC}=0$.

\section{Conclusions}
\label{sec:conclusions}

In summary, we investigate several quantum information theoretic properties of analog of the entanglement negativity in fermionic systems. This quantity was recently introduced~\cite{Shap_pTR} as a measure of entanglement in mixed states of fermions based on the fermionic partial transpose of the density matrix. Using this analogy, we have called it fermionic entanglement negativity. Among various properties an entanglement measure should satisfy, we show that the fermionic entanglement negativity is non-increasing (monotone) under LOCCs which preserve the fermion-number parity. Furthermore, the fermionic negativity is additive when the Hilbert space is enlarged by adding extra degrees of freedom through a tensor product and invariant under local unitary operations.

We further discuss the relation between the separability criterion and entanglement negativity in fermionic systems. As expected (and similar to qubit systems), the fermionic negativity of a separable state is identically zero. 

It is worth noting that the transformation rule for the \emph{fermionic} partial transpose in the Fock space (see Eq.~(\ref{eq:app_f_21})) contains a phase factor in addition to the \emph{bosonic} partial transpose (i.e., matrix transposition). In the case where each subsystem has the same fermion-number parity on both sides (inside bra and ket) the phase factor is identically one and fermionic partial transpose becomes a matrix transposition. 
This property is independent of the representation and in general we may say:
\begin{rmk}
For a bipartite state density matrix $\rho  \in {\cal S}(\Hi^A \otimes \Hi^B)$ with even fermion-number parity in each subsystem, i.e.,
\begin{align}
[(-1)^{F_{A}},\rho]= 0,
\end{align}
the fermionic and bosonic entanglement negativities are identical.
\end{rmk}


This in turn leads us to divide the density matrices of fermions into two categories.

\begin{rmk}
There exist two types of states in fermionic systems:

\begin{enumerate}[I.]
\item Fermion-number parity of subsystems is even, 
\begin{align}
[(-1)^{F_{A}},\rho]= 0,
\end{align}
\item Fermion-number parity of subsystems is mixed, 
\begin{align} \label{eq:insep_II}
[(-1)^{F_{A}},\rho]\neq 0.
\end{align}
\end{enumerate}
\end{rmk}
It is worth noting that there is no density matrix with only terms of odd subsystem fermion-number parity, since $\rho$ must always contain terms with even subsystem parity to guarantee $\tr(\rho) =1$.
 It is easy to deduce from the definition of a separable state (\ref{eq:separability}) that the first type of states can be either separable or inseparable, whereas the second type is always inseparable. As mentioned, for the first type of states, there is no difference between fermionic and bosonic entanglement negativities and we expect that all the known results about entanglement and separability in the context of  qubit systems hold in this case as well.
However, type II density matrices are specific to fermionic systems. A canonical example of this case is when we consider bipartitioning a Hilbert space of dimension $2^{N}$ into a single fermionic mode and the rest (i.e., the Hilbert space is decomposed as $2\times 2^{N-1}$ as we did in Sec.~\ref{sec:thm2}). Any inseparable state in this bipartite system is necessarily type II. We prove that the entanglement negativity of such states is positive.
We believe that this observation could be generalized to the following conjecture.


\begin{conj}
The entanglement negativity of inseparable states of type II (\ref{eq:insep_II}) is always non-vanishing,
\begin{align}
\label{eq:conj1}
{\cal E}(\rho)>0.
\end{align}
\end{conj}



An immediate consequence of this conjecture is that vanishing fermionic entanglement negativity is a necessary and sufficient condition for separability, as long as type I inseparable states are excluded. 
This property is particularly relevant to fermionic systems realized in condensed matter setups, where one wants to study the entanglement in the ground state or finite-temperature state of a Hamiltonian. Having type I inseparable states as a ground (finite-temperature) state of a fermionic Hamiltonian requires fine-tuned models which contain only terms that preserve the subsystem fermion-number parity. Therefore, as far as a generic Hamiltonian of fermions possibly with  hopping terms, pairing terms, and interactions is concerned, non-zero negativity implies an inseparable state.



In order to show the versatility of the fermionic entanglement negativity, we use it to classify entangled states in systems with small Hilbert spaces containing two or three fermionic modes.
As we have learned, the fermionic entanglement negativity is a faithful and intrinsic measure of entanglement in such systems; hence,
it would be interesting to see applications of the fermionic negativity in characterizing various quantum processes and states for measurement-based fermionic quantum computation~\cite{Bravyi2002,Chiu2013,Xin2014}.

The entanglement negativity of finite-temperature systems is another direction for research. It is well-known that the entanglement of fermions with a Fermi surface obeys a logarithmic scaling and violates the area law in all dimensions. In a future work, we would like to address how thermal fluctuations affect the logarithmic scaling behavior~\cite{Shap_finT}.
It is also worth looking at the finite-temperature states
of topologically ordered phases of fermions in two dimensions, e.g. fractional quantum Hall systems. The 2D (non-chiral) topological order is known to be unstable against thermal fluctuations~\cite{Dennis2002}. It would be interesting to see if the entanglement negativity indicates some signatures of this instability~\cite{Castelnovo2007,*Castelnovo2008,Castelnovo2017}.


Last but not least, it would be nice to study the out-of-equilibrium dynamics through the window of entanglement negativity and pinpoint possible differences between integrable and non-integrable (ergodic) systems~\cite{Gullans2018}.

\acknowledgments

We would like to thank K.~Shiozaki, T.H.~Hsieh , M.-T.~Tan, F.~Setiawan, P.-Y.~Chang, X.~Wen, and J.~Eisert for insightful discussions.
We especially thank Z.~Zimbor\'as for valuable discussions and helpful comments on the manuscript.
We are grateful to the KITP Program
{\it  Quantum Physics of Information}
(Sep 18 - Dec 15, 2017), where some part of the work was performed.
H.S. acknowledges the financial support from the KITP graduate fellowship program. 
This research was supported in part by the National Science Foundation
under Grants No.\ DMR-1455296,
and 
 No.\ NSF PHY-1748958.



\appendix

\section{Proof of Theorem~\ref{thm:2f_sep}}
\label{sec:2f_proof}

Here, we show that for a two-fermion system if ${\cal N}(\rho)=0$ then $\rho$ is separable.
In what follows, we approach this problem by proving the equivalent statement that if $\rho$ is inseparable then ${\cal N}(\rho)>0$. For concreteness, let us consider two fermions denoted by $f_1^\dag$ and $f_2^\dag$,
where the density matrix is $4\times4$ and can be represented in the basis $\{\ket{0}, f^{\dag}_1 \ket{0}, f^{\dag}_2 \ket{0}, f^{\dag}_1 f^{\dag}_2 \ket{0} \}$. A generic density matrix of fermions obeys the form
\begin{align}
\label{eq:4x4matrix}
\rho =
\begin{pmatrix}
\times &  0 & 0 &\times   \\
0 & \times  & \times  & 0 \\
0 & \times  & \times  & 0 \\
\times  & 0 & 0 & \times  \\
\end{pmatrix} 
\end{align}
where non-zero entries are shown as $\times$. 
This form of density matrix is enforced by our original assumption that only even fermion-number parity entries are allowed for the reduced density matrix of fermion-number parity symmetric systems. As we discuss in the main text, a separable state only contains diagonal elements for a two-fermion system. 
The most general form of a separable state can be written as
\begin{align}
\rho_{s}=  \frac{\mathbb{I}}{4}  + a_1 \sigma^z_A\otimes \mathbb{I}_B+ a_2  \mathbb{I}_A\otimes \sigma^z_B+
a_3 \sigma^z_A\otimes \sigma^z_B,
\end{align}
where $a_i$ are real numbers.
The eigenvalues of the this density matrix are given by
\begin{align}
\lambda_{1,2} &= \frac{1}{4}-a_3 \pm (a_1-a_2), \\
\lambda_{3,4} &= \frac{1}{4}+a_3 \pm (a_1+a_2).
\end{align}
The fact that $\rho$ is positive definite in turn implies that $|a_i|\leq 1/4$. It is easy to see that the partial transpose does nothing to the density matrix and we have $\rho_s^{\sT_{A}}=\rho_s$, i.e.,~${\cal N}(\rho_s)=0$.

Now, suppose the density matrix $\rho$ is inseparable which means that $\rho$ contains off-diagonal elements.
In general, a $4\times4$ Hermitian matrix  of the form (\ref{eq:4x4matrix}) can be represented by
\begin{align}
\rho= \rho_s + \sum_{\alpha,\beta=x,y} b_{\alpha\beta} \sigma^\alpha_A\otimes \sigma^\beta_B,
\end{align}
where $b_{\alpha\beta}$ are real coefficients.
The partial transpose is thus given by
\begin{align}
\rho^{\sT_{A}}= \rho_s +i \sum_{\alpha,\beta=x,y} c_{\alpha\beta} \sigma^\alpha_A\otimes \sigma^\beta_B,
\end{align}
in which the transformed coefficients are
\begin{align}
c_{x\beta}=b_{x\beta}, \qquad c_{y\beta}=-b_{y\beta},
\end{align}
as a result of the following rules for the partial transpose,
\begin{align}
(\sigma_A^x)^{\sT_{A}} &= i \sigma_A^x, \quad  (\sigma_A^y)^{\sT_{A}}=- i \sigma_A^y, \quad 
(\sigma_A^z)^{\sT_{A}}=  \sigma_A^z \\
(\sigma_B^\beta)^{\sT_{A}}&= \sigma_B^\beta, \quad \beta=x,y,z.
\end{align}
The trace norm is computed in terms of square root of eigenvalues of the matrix $\rho^{\sT_{A}} (\rho^{\sT_{A}})^\dag$ which is found by
\begin{align}
\rho^{\sT_{A}} (\rho^{\sT_{A}})^\dag =& \rho_s^2 + \left(\sum_{\alpha,\beta=x,y} c_{\alpha\beta} \sigma^\alpha_A\otimes \sigma^\beta_B\right)^2 \nonumber \\
&- i \left[\rho_s,\sum_{\alpha,\beta=x,y} c_{\alpha\beta} \sigma^\alpha_A\otimes \sigma^\beta_B\right].
\end{align}
After a little bit of algebra, the eigenvalues of $\sqrt{\rho^{\sT_{A}} (\rho^{\sT_{A}})^\dag}$ are found to be
\begin{align}
\lambda_{1,2} =&\left( (c_{xx}+c_{yy})^2+(c_{xy}-c_{yx})^2+(\frac{1}{4}-a_3)^2\right)^{1/2}
\nonumber \\ &\pm (a_1-a_2) \\
\lambda_{3,4} =&\left( (c_{xx}-c_{yy})^2+(c_{xy}+c_{yx})^2+(\frac{1}{4}+a_3)^2\right)^{1/2}
\nonumber \\ &\pm (a_1+a_2).
\end{align}
We should note that the above singular values of $\rho^{\sT_{A}}$ are strictly larger than those of $\rho_s^{\sT_{A}}=\rho_s$. Hence, the trace norm is greater than one, simply because
\begin{align}
\Tr |\rho^{\sT_{A}}| =& \sum_{i=1}^4 \lambda_i \nonumber \\
=& 2 \left( (c_{xx}+c_{yy})^2+(c_{xy}-c_{yx})^2+(\frac{1}{4}-a_3)^2\right)^{1/2} \nonumber \\
&+2 \left( (c_{xx}-c_{yy})^2+(c_{xy}+c_{yx})^2+(\frac{1}{4}+a_3)^2\right)^{1/2} 
\label{eq:neg_generic_2cf}
\\
\geq& 2 \left(\frac{1}{4}-a_3\right)+2 \left(\frac{1}{4}+a_3\right)=1.
\end{align}
 The inequality $\Tr|\rho^{\sT_{A}}|\geq 1$  implies that the logarithmic negativity is always positive for an inseparable state (which necessarily have off-diagonal elements in the density matrix).

\section{Proof of useful identities}
\label{sec:useful identities}

In this appendix, we provide proof of various identities discussed in the main text. 
We use a mathematical style (Proposition, Example, etc) to highlight our key results.
We start with two basic propositions.

\begin{prop}
For a local physical operator $X_B \in {\cal G}_0(\mathcal{H}^B)$ and a density matrix $\rho  \in {\cal S}( \mathcal{H}^A \otimes \mathcal{H}^B)$, we have
\begin{align}
\label{eqapp:BrT_1}
[\rho (\mathbb{I}_A\otimes X_B)]^{\sT_A}= \rho^{\sT_A} (\mathbb{I}_A\otimes X_B), 
\end{align}
or
\begin{align}
\label{eqapp:BrT_2}
[ (\mathbb{I}_A\otimes X_B) \rho]^{\sT_A}=  (\mathbb{I}_A\otimes X_B) \rho^{\sT_A}  , 
\end{align}
where $\mathbb{I}_A$ is the identity operator in $\mathcal{H}^A$.
\end{prop}

\noindent Proof:

The identities (\ref{eqapp:BrT_1}) and (\ref{eqapp:BrT_2}) follow immediately from the definition (\ref{eq:fermion_pt}).

\begin{prop}
For a local physical operator $X_A \in {\cal G}_0(\mathcal{H}^A)$ and a density matrix $\rho  \in {\cal S}(\mathcal{H}^A \otimes \mathcal{H}^B)$, we have
\begin{align}
\label{eqapp:ArT_1}
[\rho (X_A\otimes \mathbb{I}_B)]^{\sT_A}= (X_A^{\sT}\otimes \mathbb{I}_B) \rho^{\sT_A}, 
\end{align}
or
\begin{align}
\label{eqapp:ArT_2}
[(X_A\otimes \mathbb{I}_B) \rho]^{\sT_A}= \rho^{\sT_A} (X_A^{\sT}\otimes \mathbb{I}_B) , 
\end{align}
where $\mathbb{I}_B$ is the identity operator in $\mathcal{H}^B$.
\end{prop}

\noindent Proof:

Let us consider expansion of the operators in terms of Majorana operators
\begin{align}
\rho &= \sum_{k_1,k_2}^{k_1+k_2 = {\rm even}} \rho_{p_1 \cdots p_{k_1}, q_1 \cdots q_{k_2}} a_{p_1} \cdots a_{p_{k_1}} b_{q_1} \cdots b_{q_{k_2}}, \\
X_A &= \sum_{k_3= {\rm even}} X_{A,s_1 \cdots s_{k_3}} a_{s_1} \cdots a_{s_{k_3}}, 
\end{align}
There are two types of terms: Terms in which there is no Majorana operator in common between $A$ and $\rho$ and terms in which there are some common operators. In what follows, we show that (\ref{eqapp:ArT_1}) holds in both cases. A similar proof can be given for (\ref{eqapp:ArT_2}).

\begin{widetext}
Let us take the first type of terms where no Majorana operator is in common. We start from the LHS of (\ref{eqapp:ArT_1}) and arrive at the RHS.
\begin{align}
[\rho (X_A\otimes \mathbb{I}_B)]^{\sT_A}:&\nonumber \\
[(a_{p_1} \cdots a_{p_{k_1}} b_{q_1} \cdots b_{q_{k_2}})
( a_{s_1} \cdots a_{s_{k_3}}) ]^{\sT_A} &=
[( a_{s_1} \cdots a_{s_{k_3}})(a_{p_1} \cdots a_{p_{k_1}} b_{q_1} \cdots b_{q_{k_2}}) ]^{\sT_A} 
\\
&=i^{k_1+k_3}( a_{s_1} \cdots a_{s_{k_3}}) (a_{p_1} \cdots a_{p_{k_1}} b_{q_1} \cdots b_{q_{k_2}})
 \\
&= (i^{k_3} a_{s_1} \cdots a_{s_{k_3}})   (i^{k_1} a_{p_1} \cdots a_{p_{k_1}} b_{q_1} \cdots b_{q_{k_2}}) \\
&= (a_{s_1} \cdots a_{s_{k_3}})^{\sT}   ( a_{p_1} \cdots a_{p_{k_1}} b_{q_1} \cdots b_{q_{k_2}})^{\sT_A} \\
& : (A^{\sT}\otimes \mathbb{I}_2) \rho^{\sT_A}\nonumber
\end{align}
In the first line we use the fact that $k_3$ and $k_1+k_2$ are even.

We now consider the terms where some Majorana operators are in common. 
As a warm-up exercise, we start with a term where there is only one Majorana operator in common. For instance, when $a_{s_{j+1}}=a_{p_{i+1}}$. In the following, we compute the partial transpose of LHS and RHS of (\ref{eqapp:ArT_1}) and show that they are identical.
\begin{align}
 \rho (X_A\otimes \mathbb{I}_B) : &\nonumber \\
 &(a_{p_1} \cdots a_{p_i} a_{p_{i+1}}a_{p_{i+2}} \cdots a_{p_{k_1}} b_{q_1} \cdots b_{q_{k_2}})
( a_{s_1}\cdots a_{s_j} a_{s_{j+1}}a_{s_{j+2}} \cdots a_{s_{k_3}})  \\
\label{eq:1comm_l1}
=&  (a_{s_1}\cdots a_{s_j}) (a_{p_1} \cdots a_{p_i} a_{p_{i+1}}a_{p_{i+2}} \cdots a_{p_{k_1}} b_{q_1} \cdots b_{q_{k_2}})
(  a_{s_{j+1}}a_{s_{j+2}} \cdots a_{s_{k_3}}) \\
=& (-1)^{(k_1+k_2-i-1)} (a_{s_1}\cdots a_{s_j}) (a_{p_1} \cdots a_{p_i} a_{p_{i+1}}a_{s_{j+1}} a_{p_{i+2}} \cdots a_{p_{k_1}} b_{q_1} \cdots b_{q_{k_2}})
(  a_{s_{j+2}} \cdots a_{s_{k_3}}) \\
\label{eq:1comm_l3}
=& (-1)^{(i+1)+(k_3-j-1)(k_1+k_2-1)} (a_{s_1}\cdots a_{s_j} a_{s_{j+2}} \cdots a_{s_{k_3}}) (a_{p_1} \cdots a_{p_i} a_{p_{i+2}} \cdots a_{p_{k_1}} b_{q_1} \cdots b_{q_{k_2}}) \\
\label{eq:1comm_l4}
=& (-1)^{i-j} (a_{s_1}\cdots a_{s_j} a_{s_{j+2}} \cdots a_{s_{k_3}}) (a_{p_1} \cdots a_{p_i} a_{p_{i+2}} \cdots a_{p_{k_1}} b_{q_1} \cdots b_{q_{k_2}}) 
\end{align}
In (\ref{eq:1comm_l1}), we use the fact that $k_1+k_2$ is even. In going from (\ref{eq:1comm_l3}) to (\ref{eq:1comm_l4}), we utilize the fact that both $k_1+k_2$ and $k_3$ are even.
The partial transpose of the above term is then simplified into
\begin{align}
[\rho (X_A\otimes \mathbb{I}_B)]^{\sT_A}: i^{k_1+k_3-2} (-1)^{i-j} (a_{s_1}\cdots a_{s_j} a_{s_{j+2}} \cdots a_{s_{k_3}}) (a_{p_1} \cdots a_{p_i} a_{p_{i+2}} \cdots a_{p_{k_1}} b_{q_1} \cdots b_{q_{k_2}}).
\end{align}
For the RHS of (\ref{eqapp:ArT_1}), we can write
\begin{align}
(X_A^{\sT}\otimes \mathbb{I}_B) \rho^{\sT_A}: & \nonumber \\ 
&(i^{k_3} a_{s_1}\cdots a_{s_j} a_{s_{j+1}}a_{s_{j+2}} \cdots a_{s_{k_3}}) 
(i^{k_1} a_{p_1} \cdots a_{p_i} a_{p_{i+1}}a_{p_{i+2}} \cdots a_{p_{k_1}} b_{q_1} \cdots b_{q_{k_2}})\\
=&  i^{k_1+k_3} (-1)^{k_3-j-1+i} (a_{s_1}\cdots a_{s_j} a_{s_{j+2}} \cdots a_{s_{k_3}}) 
(a_{p_1} \cdots a_{p_i} a_{s_{j+1}} a_{p_{i+1}}a_{p_{i+2}} \cdots a_{p_{k_1}} b_{q_1} \cdots b_{q_{k_2}})\\
=&  i^{k_1+k_3} (-1)^{i-j-1} (a_{s_1}\cdots a_{s_j} a_{s_{j+2}} \cdots a_{s_{k_3}}) 
(a_{p_1} \cdots a_{p_i} a_{p_{i+2}} \cdots a_{p_{k_1}} b_{q_1} \cdots b_{q_{k_2}})\\
:& [\rho (X_A\otimes \mathbb{I}_B)]^{\sT_A} \nonumber
\end{align}
and the proof of this type of terms is complete.
Let us explicitly check that the corresponding term in the wrong ordering, i.e.,  $\rho^{\sT_A} (X_A^{\sT}\otimes \mathbb{I}_B) $ is not equal to that of $ [\rho (X_A\otimes \mathbb{I}_B)]^{\sT_A} $.
\begin{align}
\rho^{\sT_A} (X_A^{\sT}\otimes \mathbb{I}_B) : &\nonumber \\ 
& (i^{k_1} a_{p_1} \cdots a_{p_i} a_{p_{i+1}}a_{p_{i+2}} \cdots a_{p_{k_1}} b_{q_1} \cdots b_{q_{k_2}}) (i^{k_3} a_{s_1}\cdots a_{s_j} a_{s_{j+1}}a_{s_{j+2}} \cdots a_{s_{k_3}}) \\
=& i^{k_1+k_3} (a_{p_1} \cdots a_{p_i} a_{p_{i+1}}a_{p_{i+2}} \cdots a_{p_{k_1}} b_{q_1} \cdots b_{q_{k_2}})
( a_{s_1}\cdots a_{s_j} a_{s_{j+1}}a_{s_{j+2}} \cdots a_{s_{k_3}})  \\
\label{eq:1commW_l3}
=&i^{k_1+k_3} (-1)^{i-j} (a_{s_1}\cdots a_{s_j} a_{s_{j+2}} \cdots a_{s_{k_3}}) (a_{p_1} \cdots a_{p_i} a_{p_{i+2}} \cdots a_{p_{k_1}} b_{q_1} \cdots b_{q_{k_2}}) \\
&: -  [\rho (X_A\otimes \mathbb{I}_B)]^{\sT_A} \nonumber
\end{align}
where in order to get (\ref{eq:1commW_l3}) we use the result in (\ref{eq:1comm_l4}).
So, there is a minus sign appearing and they are not equal.

Now, consider a term where there are $n$ Majorana operators $\{a_{p_i}\}$ for $i=1,\cdots,n$ in common between $\rho$ and $A$. 
We represent such a term in $\rho$ by
\begin{align}
\rho:  O^\rho_{p_0} a_{p_1} O^\rho_{p_1} a_{p_2} O^\rho_{p_2} \cdots a_{p_n} O^\rho_{p_{n}} 
b_{q_1} \cdots b_{q_{k_2}}
\end{align}
where the $O^\rho_{p_i}$ operator contains only $l_i$ non-repeating Majorana operators and similarly for a term in $X_A$, we write
\begin{align}
X_A:  O^X_{p_0} a_{p_1} O^X_{p_1} a_{p_2}O^X_{p_2}  \cdots a_{p_n} O^X_{p_{n}}  
\end{align}
where $O^X_{p_i}$ contains only $l'_i$ non-repeating Majorana operators. Note that $k_2+n+\sum_{i=0}^n l_i$ and  $n+\sum_{i=0}^n l_i'$ are even numbers and $n$ can be even or odd. 

Let us start with the LHS of (\ref{eqapp:ArT_1}) before taking the partial transpose
\begin{align}
 \rho (X_A\otimes \mathbb{I}_B) : &\nonumber \\
& (O^\rho_{p_0} a_{p_1} O^\rho_{p_1} a_{p_2} O^\rho_{p_2}  \cdots  a_{p_n} O^\rho_{p_{n}} 
b_{q_1} \cdots b_{q_{k_2}}) (O^X_{p_0} a_{p_1} O^X_{p_1} a_{p_2} O^X_{p_2}  \cdots a_{p_n} O^X_{p_{n}})
\\
 =&
 (-1)^{k_2+k_1-l_0-1} O^X_{p_0} (O^\rho_{p_0} a_{p_1}a_{p_1} O^\rho_{p_1} a_{p_2} O^\rho_{p_2}  \cdots  a_{p_n} O^\rho_{p_{n}} 
b_{q_1} \cdots b_{q_{k_2}}) (  O^X_{p_1} a_{p_2} O^X_{p_2}  \cdots a_{p_n} O^X_{p_{n}}) \\
 =&
 (-1)^{l_0+1+l_1'(k_1+k_2-1)} O^X_{p_0} O^X_{p_1} (O^\rho_{p_0} O^\rho_{p_1} a_{p_2} O^\rho_{p_2}  \cdots  a_{p_n} O^\rho_{p_{n}} 
b_{q_1} \cdots b_{q_{k_2}}) (   a_{p_2} O^X_{p_2}  \cdots a_{p_n} O^X_{p_{n}}) \\
 =&
 (-1)^{l_0+1+l_1'+l_0+l_1+2} O^X_{p_0} O^X_{p_1} (O^\rho_{p_0} O^\rho_{p_1} a_{p_2}a_{p_2}  O^\rho_{p_2}  \cdots  a_{p_n} O^\rho_{p_{n}} 
b_{q_1} \cdots b_{q_{k_2}}) (   O^X_{p_2}  \cdots a_{p_n} O^X_{p_{n}}) 
\\
 =&
 (-1)^{l_0+1+l_1'+l_0+l_1+2+2 l_2'} O^X_{p_0} O^X_{p_1} O^X_{p_2} (O^\rho_{p_0} O^\rho_{p_1}   O^\rho_{p_2}  \cdots  a_{p_n} O^\rho_{p_{n}} 
b_{q_1} \cdots b_{q_{k_2}}) (     \cdots a_{p_n} O^X_{p_{n}}) 
\end{align}
we continue this process and we finally get
\begin{align}
 \rho (X_A\otimes \mathbb{I}_B) :&
 (-1)^{\phi_1} (O^X_{p_0} O^X_{p_1} O^X_{p_2}  \cdots O^X_{p_{n}})  (O^\rho_{p_0} O^\rho_{p_1}   O^\rho_{p_2}  \cdots O^\rho_{p_{n}} 
b_{q_1} \cdots b_{q_{k_2}}) 
\end{align}
where
\begin{align}
\phi_1= \sum_{i=0}^n [i l'_i + (n-i) l_i] + \frac{n(n+1)}{2}.
\end{align}
Hence, the LHS of (\ref{eqapp:ArT_1}) is given by
\begin{align}
 [\rho (X_A\otimes \mathbb{I}_B)]^{\sT_A} :&
(-1)^{\phi_1+\alpha} (O^X_{p_0} O^X_{p_1} O^X_{p_2}  \cdots O^X_{p_{n}})  (O^\rho_{p_0} O^\rho_{p_1}   O^\rho_{p_2}  \cdots O^\rho_{p_{n}} 
b_{q_1} \cdots b_{q_{k_2}}),
\end{align}
where
\begin{align}
\alpha= \frac{1}{2} \sum_{i=0}^n (l_i+l_i').
\end{align}
Next, we compute the RHS
\begin{align}
(X_A^{\sT}\otimes \mathbb{I}_B)\rho^{\sT_A}: & \nonumber \\
&(-1)^{\alpha+ n} (O^X_{p_0} a_{p_1}O^X_{p_1}   \cdots a_{p_{n-1}} O^X_{p_{n-1}} a_{p_n} O^X_{p_{n}})  (O^\rho_{p_0} a_{p_1} O^\rho_{p_1}   \cdots a_{p_{n-1}} O^\rho_{p_{n-1}}  a_{p_n} O^\rho_{p_{n}} 
b_{q_1} \cdots b_{q_{k_2}}) \\
=&(-1)^{\alpha+ n + l_n'+ \sum_{i=0}^{n-1} l_i + n-1} (O^X_{p_0}  a_{p_1} O^X_{p_1}   \cdots a_{p_{n-1}} O^X_{p_{n-1}} O^X_{p_{n}})  (O^\rho_{p_0} a_{p_1} O^\rho_{p_1} \cdots a_{p_{n-1}} O^\rho_{p_{n-1}} a_{p_n} a_{p_n} O^\rho_{p_{n}} 
b_{q_1} \cdots b_{q_{k_2}})\\
=& (-1)^{\alpha+ n + l_n'+ \sum_{i=0}^{n-1} l_i + n-1+ l_{n}'+ l_{n-1}'+ \sum_{i=0}^{n-2} l_i + n-2} (O^X_{p_0} a_{p_1} O^X_{p_1}    \cdots O^X_{p_{n-1}} O^X_{p_{n}})\nonumber \\
& \times  (O^\rho_{p_0} O^\rho_{p_1} O^\rho_{p_2}  \cdots a_{p_{n-1}}a_{p_{n-1}} O^\rho_{p_{n-1}} O^\rho_{p_{n}} 
b_{q_1} \cdots b_{q_{k_2}})
\end{align}
and finally we obtain
\begin{align}
(X_A^{\sT}\otimes \mathbb{I}_B)\rho^{\sT_A}:  (-1)^{\alpha+ n +\phi_2} (O^X_{p_0} O^X_{p_1} O^X_{p_2}  \cdots  O^X_{p_{n}})  (O^\rho_{p_0} O^\rho_{p_1}  O^\rho_{p_2}  \cdots  O^\rho_{p_{n}} 
b_{q_1} \cdots b_{q_{k_2}})
\end{align}
\end{widetext}
in which
\begin{align}
\phi_2= \sum_{i=0}^n [i l'_i + (n-i) l_i] + \frac{n(n-1)}{2}.
\end{align}
Evidently, phases do match since $\phi_1=\phi_2+n$ and therefore, the identity (\ref{eqapp:ArT_1}) holds.
\hfill $\blacksquare$

\noindent {\bf Proof of Proposition \ref{prop:ABrT_1} of main text:} The equality (\ref{eq:ABrT_1}) directly follows from (\ref{eqapp:BrT_1}) and (\ref{eqapp:ArT_1}) as in
\begin{align}
[\rho (X_A\otimes X_B)]^{\sT_A} &= [\rho  (X_A\otimes \mathbb{I}_B) (\mathbb{I}_A\otimes X_B)]^{\sT_A}  \\
&= [\rho  (X_A\otimes \mathbb{I}_B)]^{\sT_A} (\mathbb{I}_A\otimes X_B) \\
&= (X_A^{\sT} \otimes \mathbb{I}_B) \rho^{\sT_A} (\mathbb{I}_A\otimes X_B).
\end{align}
Equation (\ref{eq:ABrT_2}) can be proved similarly.
\hfill $\blacksquare$

We already have $(\rho^{\sT_A})^{\sT_B}=\rho^{\sT}$ for a density matrix operator from the definition (\ref{eq:fermion_pt}).
Now, let us do some consistency checks.

\begin{chk}
For a local physical operator $X_B \in {\cal G}_0(\mathcal{H}^B)$ and a density matrix $\rho  \in {\cal S}(\mathcal{H}^A \otimes \mathcal{H}^B)$, we should satisfy
\begin{align}
([\rho (\mathbb{I}_A\otimes X_B)]^{\sT_A})^{\sT_B} \stackrel{?}{=}  [\rho (\mathbb{I}_A\otimes X_B)]^{\sT}.
\end{align}
\end{chk}

\noindent Answer:
\begin{align}
([\rho (\mathbb{I}_A\otimes X_B)]^{\sT_A})^{\sT_B} &= (\rho^{\sT_A} (\mathbb{I}_A\otimes X_B))^{\sT_B} 
\\
&= (\mathbb{I}_A\otimes X_B^{\sT}) (\rho^{\sT_A})^{\sT_B} \\
&= (\mathbb{I}_A\otimes X_B^{\sT}) \rho^{\sT} \\
&=[\rho (\mathbb{I}_A\otimes X_B)]^{\sT}.
\end{align}
First and second lines follow from (\ref{eqapp:BrT_1}) and (\ref{eqapp:ArT_1}), respectively.

\begin{chk}
For a local physical operator $X_A \in {\cal G}_0(\mathcal{H}^A)$ and $\rho  \in {\cal S}(\mathcal{H}^A \otimes \mathcal{H}^B)$, we should satisfy
\begin{align}
([\rho (X_A\otimes \mathbb{I}_B)]^{\sT_A})^{\sT_B} \stackrel{?}{=}   [\rho  (X_A\otimes \mathbb{I}_B)]^{\sT}.
\end{align}
\end{chk}

\noindent Answer:
\begin{align}
([\rho  (X_A\otimes \mathbb{I}_2)]^{\sT_{A}})^{\sT_{B}} &= ( (X_A^{\sT}\otimes \mathbb{I}_B) \rho^{\sT_A})^{\sT_B} 
\\
&=  (X_A^{\sT}\otimes \mathbb{I}_B)  (\rho^{\sT_A})^{\sT_B} \\
&= (X_A^{\sT}\otimes \mathbb{I}_B)  \rho^{\sT} \\
&=  [\rho  (X_A\otimes \mathbb{I}_B)]^{\sT}.
\end{align}
First and second lines follow from (\ref{eqapp:ArT_1}) and  (\ref{eqapp:BrT_2}), respectively.

\begin{chk}
For two local physical operators $X_A \in {\cal G}_0(\mathcal{H}^A)$, $X_B \in {\cal G}_0(\mathcal{H}^B)$ and a density matrix
$\rho  \in {\cal S}(\mathcal{H}^A \otimes \mathcal{H}^B)$, we need to satisfy
\begin{align}
([\rho (X_A\otimes X_B)]^{\sT_A})^{\sT_B}  \stackrel{?}{=} [\rho (X_A\otimes X_B)]^{\sT} .
\end{align}
\end{chk}

\noindent Answer:
\begin{align}
([\rho (X_A\otimes X_B)]^{\sT_A})^{\sT_B} &= ((X_A^{\sT} \otimes \mathbb{I}_B) [\rho^{\sT_A} (\mathbb{I}_A\otimes X_B)])^{\sT_B}  \\
 &= (X_A^{\sT} \otimes \mathbb{I}_B) [\rho^{\sT_A} (\mathbb{I}_A\otimes X_B)]^{\sT_B}  \\
&= (X_A^{\sT} \otimes \mathbb{I}_B) (\mathbb{I}_A\otimes X_B^{\sT}) (\rho^{\sT_A})^{\sT_B} \\
&= (X_A^{\sT} \otimes X_B^{\sT}) \rho^{\sT}\\
&= [\rho (X_A\otimes X_B)]^{\sT}.
\end{align}

\begin{chk}
For local physical operators $X_A,Y_A \in {\cal G}_0(\mathcal{H}^A)$,
$X_B,Y_B \in {\cal G}_0(\mathcal{H}^B)$ and $\rho  \in {\cal S}(\mathcal{H}^A \otimes \mathcal{H}^B)$, we have to satisfy
\begin{align}
([ (X_A\otimes X_B) \rho (Y_A \otimes Y_B)]^{\sT_A})^{\sT_B} \stackrel{?}{=} [ (X_A\otimes X_B) \rho (Y_A\otimes Y_B)]^{\sT}
\end{align}
\end{chk}

\noindent Answer:
\begin{align}
&([ (X_A\otimes X_B) \rho (Y_A\otimes Y_B)]^{\sT_A})^{\sT_B}\nonumber  \\
&= 
((Y_A^{\sT} \otimes B) \rho^{\sT_A} (A^{\sT} \otimes Y_B))^{\sT_B} \\
&= (Y_A^{\sT} \otimes Y_B^{\sT}) (\rho^{\sT_A})^{\sT_B} (A^{\sT} \otimes B^{\sT}) \\
&= (Y_A^{\sT} \otimes Y_B^{\sT}) \rho^{\sT} (A^{\sT} \otimes B^{\sT})\\
&=[ (A\otimes B) \rho (Y_A\otimes Y_B)]^{\sT}.
\end{align}

\section{Proof of Theorem~\ref{thm:bisep}}
\label{sec:perturb_proof}

Here, we show that the negativity is non-zero right outside the boundary of the set of separable states. The inseparable states in the vicinity of separable states is obtained by adding an infinitesimal off-diagonal term to a separable density matrix. We use perturbation theory to show that this infinitesimal term gives a positive contribution to the negativity.

We begin by noting that a general form of a separable state can then be written as
\begin{align}
\rho_{\text{sep}}= w_0 \vac \otimes \rho_0 + w_1 f_1^\dag \vac f_1 \otimes \rho_1 
\end{align}
where
$f_1\in {\cal G}(\Hi^A)$, $\rho_1, \rho_2 \in {\cal S}(\Hi^{\bar A})$ are two fermionic density matrices and $w_0+w_1=1$. 
 An inseparable state can be written in the form 
\begin{align}
\rho=\rho_{\text{sep}} + \rho_{\text{off}}
\end{align}
where
\begin{align}
\rho_{\text{off}} =  f_1^\dag \vac  \otimes \delta \rho +  \vac f_1   \otimes \delta \rho^\dag,
\end{align}
where $\delta\rho \in {\cal G}_1(\Hi^{\bar A}) $ is a fermion-number parity odd operator. The partial transpose is given by
\begin{align}
\rho^{\sT_A}=\rho_{\text{sep}} + i  f_1^\dag \vac  \otimes \delta \rho^\dag+ i \vac f_1   \otimes \delta \rho,
\end{align}
and its adjoint is
\begin{align}
(\rho^{\sT_A})^\dag=\rho_{\text{sep}} - i  f_1^\dag \vac  \otimes \delta \rho^\dag - i \vac f_1   \otimes \delta \rho.
\end{align}
Thus, we write 
\begin{align}
\rho^{\sT_A} (\rho^{\sT_A})^\dag= \rho^{(0)} + \delta V
\end{align}
where   
\begin{align}
\rho^{(0)}=\rho_{\text{sep}}^2 = w_0^2 \vac \otimes \rho_0^2 +  w_1^2 f_1^\dag \vac f_1 \otimes  \rho_1^2
\end{align}
and
\begin{align}
\delta V=& \vac \otimes \delta\rho \delta\rho^\dag + f_1^\dag \vac f_1 \otimes \delta\rho^\dag \delta\rho
\nonumber  \\ &+
 i  f_1^\dag \vac  \otimes (w_0  \delta \rho^\dag \rho_0-w_1\rho_1 \delta \rho^\dag) 
 \nonumber  \\ &+
  i \vac f_1   \otimes ( w_1\delta \rho \rho_1 -w_0 \rho_0 \delta \rho).
\end{align}
Now, we want to do perturbation theory and show that $\tr\sqrt{\rho^{\sT_A} (\rho^{\sT_A})^\dag}$ is larger than $\tr\sqrt{\rho^{(0)}}=\tr\rho_{\text{sep}}=1$ to the lowest order in powers of $\delta\rho$. 

Suppose $\Hi^{\bar A}$ contains $m$ fermionic modes. Let $\ket{\psi_{j}}$, and $\ket{\phi_{j}}$, $j=1,\cdots,2^m$ be the $2^m$ eigenstates of $\rho_0$ and $\rho_1$
\begin{align}
\rho_0 \ket{\psi_j}= \mu_j \ket{\psi_j}, \nonumber \\
\rho_1 \ket{\phi_j}= \nu_j \ket{\phi_j}, 
\end{align}
 with eigenvalues $\mu_j$ and $\nu_j$, such that $\sum_j \mu_j =\sum_j \nu_j=1$. It is important to note that each of $\{\ket{\psi_{j}} \}$ and $\{ \ket{\phi_{j}} \}$ forms a complete orthonormal basis in the Hilbert space $\Hi^B$.
 Hence, the unperturbed part $\rho^{(0)}$ is diagonal in this eigenbasis $\{\ket{n^{(0)}},\lambda_n^{(0)} \}$
 \begin{align*}
 \ket{n^{(0)}}=& \ket{0}\otimes \ket{\psi_j} \qquad
  \lambda_n^{(0)} = w_0^2 \mu_j^2\qquad  1\leq n \leq 2^m \\
 =& f_1^\dag \ket{0}\otimes \ket{\phi_j}  \quad
   \lambda_n^{(0)} = w_1^2 \nu_j^2 \qquad  2^m < n \leq 2^{m+1}
 \end{align*}
  The matrix elements of the perturbation term $\delta V$ is found by
\begin{align}
 &\bra{0}\otimes \bra{\psi_j} \delta V f_1^\dag \ket{0} \otimes \ket{\phi_k} \nonumber \\
 = &
 i \bra{\psi_j} ( w_1\delta \rho \rho_1 -w_0 \rho_0 \delta \rho)  \ket{\phi_k} \nonumber \\
 =&  i (w_1\nu_k- w_0 \mu_j) \bra{\psi_j} \delta \rho  \ket{\phi_k}
\end{align}
\vspace{-.5cm}
\begin{align}
 \bra{0}\otimes \bra{\psi_j} \delta V  \ket{0} \otimes \ket{\psi_k} &=\bra{\psi_j} \delta\rho\delta\rho^\dag \ket{\psi_k}  \\
 \bra{0}f_1 \otimes \bra{\phi_j} \delta V f_1^\dag \ket{0} \otimes \ket{\phi_k} &=  \bra{\phi_j} \delta\rho^\dag \delta\rho \ket{\phi_k} 
\end{align}
According to perturbation theory, corrections to an eigenvalue $\lambda_n$ of $\rho^{\sT_A} (\rho^{\sT_A})^\dag$ up to second order in $\delta\rho$ are given by
\begin{align}
\lambda_n =& \lambda_n^{(0)}+  \bra{n^{(0)}} \delta V \ket{n^{(0)}} \nonumber  \\ &+ \sum_{k\neq n} \frac{|\bra{k^{(0)}}\delta V\ket{n^{(0)}}|^2}{\lambda_n^{(0)}-\lambda_k^{(0)}} + O (\delta \rho^3),
\end{align}
where $\lambda_n^{(0)}$ is an eigenstate of unperturbed  (separable) density matrix $\rho_{\text{sep}}$.
So, we find that for $\lambda_n,\ n\leq 2^m$,
\begin{align}
\lambda_n =& w_0^2 \mu_j^2 + \bra{\psi_j} \delta\rho\delta\rho^\dag \ket{\psi_j} \nonumber  \\ &+
 \sum_{k} \left( \frac{w_0 \mu_j - w_1 \nu_k }{w_0 \mu_j + w_1 \nu_k }\right) | \bra{\psi_j} \delta \rho  \ket{\phi_k}|^2 + O (\delta \rho^3) \\
 =& w_0^2 \mu_j^2 + \sum_k \bra{\psi_j} \delta\rho \ket{\phi_k} \bra{\phi_k} \delta\rho^\dag \ket{\psi_j} \nonumber  \\ &+
 \sum_{k} \left( \frac{w_0 \mu_j - w_1 \nu_k }{w_0 \mu_j + w_1 \nu_k }\right) | \bra{\psi_j} \delta \rho  \ket{\phi_k}|^2 + O (\delta \rho^3)
 \label{eq:eigsepthm2_l2}
   \\
 =&  w_0^2 \mu_j^2  +
 2 w_0 \mu_j \sum_{k} \frac{| \bra{\psi_j} \delta \rho  \ket{\phi_k}|^2 }{w_0 \mu_j + w_1 \nu_k } + O (\delta \rho^3), 
\end{align}
and similarly for $\lambda_n,\ n> 2^m$,
 \begin{align}
\lambda_n &=  w_1^2 \nu_j^2  +
 2 w_1 \nu_j  \sum_{k} \frac{| \bra{\psi_k} \delta \rho  \ket{\phi_j}|^2}{w_1 \nu_j + w_0 \mu_k }   + O (\delta \rho^3).
\end{align}
Notice that in the first line of (\ref{eq:eigsepthm2_l2}), we inserted the resolution of identity $\sum_k \ket{\phi_k} \bra{\phi_k}=\mathbb{I}$.
Hence, the trace norm is given by
\begin{align}
\norm{\rho^{\sT_A}} &= \sum_n \sqrt{\lambda_n} = 1+2 \sum_{jk}  \frac{| \bra{\psi_j} \delta \rho  \ket{\phi_k}|^2 }{w_0 \mu_j + w_1 \nu_k }  + O (\delta \rho^3 ),
\end{align}
which implies ${\cal N}(\rho) > 0$ to this order of approximation.
It is important to note that such a contribution is absent in the bosonic partial transpose.

\section{Some properties of bosonic partial transpose of density matrices in fermionic systems}
\label{sec:bosonic_transpose}

The bosonic partial transpose $\rho^{\widetilde{\sT}_{\sA}}$ is related to the fermionic partial transpose $\rho^{\sT_A}$ as follows,
\begin{align} \label{eq:b_ptrans}
\rho^{\widetilde{\sT}_{\sA}}= \left(\frac{1+i}{2}\right) \rho^{\sT_A} + \left(\frac{1-i}{2}\right) \rho^{\sT_A\dag}
\end{align}
This relation can be easily seen in the expansion of density matrix in terms of Majorana operators~(\ref{eq:fermion_pt}). Given the properties of fermionic partial transpose, we are going to use the above relation to investigate we show that the bosonic partial transpose satisfies all of them except for the additivity.

\textbf{Separable states.$-$} From the definition (\ref{eq:separability}), we write
\begin{align}
\rho_{\text{sep}}^{\widetilde{\sT}_{\sA}} &= \sum_i w_i \left[ \left(\frac{1+i}{2}\right)  \rho_{A,i}^{\sT}  \otimes \rho_{B,i} + \left(\frac{1-i}{2}\right) ( \rho_{A,i}^{\sT}  \otimes \rho_{B,i})^\dag \right]  \nonumber \\
&= \sum_i w_i \left[ \left(\frac{1+i}{2}\right)  \rho_{A,i}^{\sT}  \otimes \rho_{B,i} + \left(\frac{1-i}{2}\right) \rho_{A,i}^{\sT}  \otimes \rho_{B,i} \right] \nonumber \\
&= \sum_i w_i \rho_{A,i}^{\sT}  \otimes \rho_{B,i}  = \rho_{\text{sep}}^{\sT_A}
\end{align}
where we use the Hermiticity of density matrices and the fact that the full transpose and hermitian conjugate commute. Hence, bosonic partial transpose also vanishes for a separable state of fermions.

\textbf{Local unitaries.$-$}
The bosonic partial transpose is invariant under application of local unitary operators, as can be seen below.
\begin{align}
 & \left[ (U_A\otimes U_B) \rho (U_A^\dag \otimes  U_B^\dag) \right]^{\widetilde{\sT}_{\sA}} \nonumber \\
=&  \left(\frac{1-i}{2}\right) ((U_A^\dag)^{\sT} \otimes U_B) \rho^{\sT_A}  (U_A^{\sT}\otimes U_B^\dag) 
\nonumber \\
&+ \left(\frac{1+i}{2}\right) ((U_A^\dag)^{\sT}\otimes U_B) \rho^{\sT_A\dag} (U_A^{\sT} \otimes U_B^\dag ) \nonumber \\
 =&    ((U_A^\dag)^{\sT} \otimes U_B) \rho^{\widetilde{\sT}_{\sA}} ( U_A^{\sT}\otimes  U_B^\dag).
\end{align}

\textbf{Appending ancilla.$-$}
Analog of (\ref{eq:anc_T}) holds for the bosonic partial transpose,
\begin{align}
  ( \rho_{AB} \otimes \rho_R)^{\widetilde{\sT}_{\tilde{\sA}}}
  &= \left(\frac{1+i}{2}\right) \rho_{AB}^{{\sT}_{\sA}} \otimes \rho_R^{\sT} + \left(\frac{1-i}{2}\right) (\rho_{AB}^{{\sT}_{\sA}} \otimes \rho_R^{\sT})^\dag
\nonumber \\
  &= \left[ \left(\frac{1+i}{2}\right) \rho_{AB}^{{\sT}_{\sA}}  + \left(\frac{1-i}{2}\right) \rho_{AB}^{{\sT}_{\sA}\dag} \right] \otimes \rho_R^{\sT} 
\nonumber \\
&=\rho_{AB}^{\widetilde{\sT}_{\sA}} \otimes \rho_R^{\sT},
\end{align}
which implies that the bosonic negativity does not change upon adding an ancilla.

\textbf{Local projectors.$-$}
Here, we show that the expression (\ref{eq:ABCDrT_1}) also holds for the bosonic partial transpose. The rest of proof follows from the same steps as those in (\ref{eq:mono3_norm}), 
\begin{align}
 & \left[ (P_A\otimes P_B) \rho (P_A \otimes  P_B) \right]^{\widetilde{\sT}_{\sA}} \nonumber \\
=&  \left(\frac{1-i}{2}\right) (P_A^{\sT} \otimes P_B) \rho^{\sT_A}  (P_A^{\sT}\otimes P_B) 
\nonumber \\
&+ \left(\frac{1+i}{2}\right) (P_A^{\sT} \otimes P_B)^\dag \rho^{\sT_A\dag} (P_A^{\sT} \otimes P_B)^\dag
\nonumber \\
 =&    (P_A^{\sT} \otimes P_B) \rho^{\widetilde{\sT}_{\sA}} (P_A^{\sT} \otimes P_B),
\end{align}
where we use the Hermiticity of the projectors.

\textbf{Tracing out ancilla.$-$}
The steps to prove the inequalities (\ref{eq:mono4_1}-\ref{eq:mono4_logN}) depend on the last three conditions above. Since the bosonic partial transpose satisfies them, it also satisfies the third monotonicity criterion.

\textbf{Additivity.$-$}
It is easy to see that why the bosonic partial transpose does not respect the tensor product structure Eq.~(\ref{eq:tensor_prod}), 
\begin{align}
(\rho_{AB} \otimes \rho_{AB}' )^{\widetilde{\sT}_{\sA}} &=
\left(\frac{1+i}{2}\right) \rho_{AB}^{\sT_A} \otimes \rho_{AB}'^{\sT_A} + \left(\frac{1-i}{2}\right) \rho_{AB}^{\sT_A\dag} \otimes \rho_{AB}'^{\sT_A\dag} \nonumber \\
&\neq  \rho_{AB}^{\widetilde{\sT}_{\sA}} \otimes \rho_{AB}'^{\widetilde{\sT}_{\sA}}.
\end{align}
The fact that the last line is not equal can be immediately understood from the definition (\ref{eq:b_ptrans}).

\bibliography{refs}

\end{document}